\documentclass[aps,prc,twoside,twocolumn,nofootinbib,10pt,floatfix]{revtex4}
\usepackage{amsmath,amssymb}
\usepackage{graphicx,bm}
\usepackage{slashed}
\usepackage{epstopdf}
\usepackage{ulem} 
\usepackage[usenames]{color}
\usepackage{float}
\usepackage{hyperref}
\usepackage{subfigure}
\usepackage{rotating}
\usepackage{color}
\usepackage{multirow}
\usepackage{dcolumn}
\usepackage{overpic}
\usepackage{booktabs}
\usepackage{makecell}
\usepackage{arydshln} 
\usepackage{diagbox}
\usepackage{tabularx}
\usepackage{array}
\newcolumntype{|}{!{\vline}}

\renewcommand\sout{\bgroup \color{red} \ULdepth=-.5ex \ULset}

\newsavebox{\tablebox}
\usepackage{adjustbox}
\begin{document}
\title{Systematic exploration of triply heavy tetraquarks: spectroscopic and decay characteristics
}
\author{Hong-Tao An$^{1}$}\email{anht@mail.tsinghua.edu.cn}
\author{Yu-Shuai Li$^{2}$}\email{liysh@pku.edu.cn}
\author{Si-Qiang Luo$^{3,4}$}\email{luosq15@lzu.edu.cn}
\affiliation{
$^1$Department of Physics and Center for High Energy Physics, Tsinghua University, Beijing 100084, China \\
$^2$ School of Physics and Center of High Energy Physics, Peking University, Beijing 100871, China\\
$^3$School of Physical Science and Technology, Lanzhou University, Lanzhou 730000, China\\
$^4$Lanzhou Center for Theoretical Physics, Key Laboratory of Theoretical Physics of Gansu Province
}
\date{\today}
\begin{abstract}
While hidden, singly, doubly, and fully heavy tetraquark states have been experimentally observed, triply heavy tetraquark states remain experimentally unconfirmed.
We systematically investigate the spectroscopic and decay properties of four triply heavy-flavor tetraquark systems ($cc\bar{c}\bar{n}$, $cc\bar{c}\bar{s}$, $bb\bar{b}\bar{n}$, $bb\bar{b}\bar{s}$; $n=u,d$) based on the nonrelativistic quark model.
Using an effective Hamiltonian, we employ the Gaussian expansion method to solve the four-body Schr\"{o}dinger equation and incorporate the effect of color-spin configuration mixing. 
Results show both $cc\bar{c}\bar{q}$ and $bb\bar{b}\bar{q}$ systems have two $J^{P}=0^{+}$, three $J^{P}=1^{+}$, and one $J^{P}=2^{+}$ states, 
with ground-state masses of 5.2–5.5 GeV and 15.0–15.3 GeV, respectively. 
Root-mean-square radius analysis supports compact tetraquark configurations. 
All states are unstable, with rearrangement strong decays dominant and negligible radiative decays. 
Narrow resonances (e.g., $T_{c^{2}\bar{c}\bar{s}}(5360,0^{+})$, $T_{b^{2}\bar{b}\bar{n}}(15148,2^{+})$) arise from Feynman amplitude cancellation. 
We propose experimental searches in $J/\psi D^{*}_{s}$/ $\eta_{c}D_{s}$ (5.3–5.4 GeV) and $\Upsilon B^{*}$ (15.1–15.2 GeV) channels, providing key guidance for triply heavy tetraquark identification.
\end{abstract}
\maketitle

\section{Introduction}\label{sec1}

Searching for multiquark states has become one of the most important research directions in hadron physics,
with experimental observations and theoretical studies continuously advancing our understanding of nonperturbative quantum chromodynamics \cite{Chen:2016qju,Guo:2017jvc,Brambilla:2019esw,Liu:2019zoy,Hosaka:2016pey,Chen:2022asf,Liu:2013waa,Berwein:2024ztx}.
Since the Belle Collaboration’s observation of X(3872) \cite{Belle:2003nnu} in 2003, 
many exotic states, including $P_{c}$ and $P_{cs}$ pentaquark states, have been experimentally observed \cite{LHCb:2015yax,LHCb:2019kea,LHCb:2020jpq,LHCb:2022ogu}.

In the sector of hidden heavy tetraquark states, besides the aforementioned X(3872),
a large number of charmonium-like and bottomonium-like states have been identified in a wide range of processes over the past two decades \cite{BESIII:2023wsc,BESIII:2013mhi,Belle:2011aa,LHCb:2022aki,CMS:2013jru,LHCb:2016axx,LHCb:2021uow}.
The internal structures of these states are widely interpreted in terms of two mainstream configurations: compact tetraquark states \cite{Anwar:2018sol,Dubnicka:2011mm,Dubnicka:2010kz,Wu:2016gas} and meson-antimeson molecular states \cite{Liu:2024ziu,Wang:2023ivd,Godfrey:2008nc,Wang:2021aql,Wang:2013kva}. 

In the sector of singly heavy tetraquark states, the BaBar Collaboration reported a narrow peak, named $D_{s0}(2317)$, in the invariant mass spectrum of $D^{+}_{s}\pi^{0}$ in 2003 \cite{BaBar:2003oey}.
Subsequently, the CLEO Collaboration observed another narrow peak, named $D_{s1}(2460)$, in the invariant mass spectrum of $D^{*+}_{s}\pi^{0}$ \cite{CLEO:2003ggt}.
These two states can alternatively be interpreted as either $c\bar{s}q\bar{q}$ tetraquark states \cite{Chen:2004dy,Cheng:2003kg,Kim:2005gt} or $DK/D^{*}K$ molecular states \cite{Guo:2006rp,Barnes:2003dj,Navarra:2015iea}.
In 2020, the LHCb Collaboration observed an exotic peak in the $D^{-}K^{+}$ channel; to fit the experimental data, the collaboration introduced two resonances,
named $T_{cs0}(2900)^{0}$ and $T_{cs1}(2900)^{0}$, with a minimal quark content of $ud\bar{s}\bar{c}$ \cite{LHCb:2020bls,LHCb:2020pxc}. 
In 2023, the LHCb Collaboration observed another two singly charmed tetraquark states, 
$T^{a}_{c\bar{s}0}(2900)^{0}$ and $T^{a}_{c\bar{s}0}(2900)^{++}$, with minimal quark contents of $u\bar{d}\bar{s}c$ and $\bar{u}d\bar{s}c$, respectively \cite{LHCb:2022sfr,LHCb:2022lzp}.  

In the sector of doubly heavy tetraquark states, the LHCb Collaboration reported the observation of a doubly charmed tetraquark candidate, $T^{+}_{cc}(3875)$, in the $D^{0}D^{0}\pi^{+}$ mass distribution just below the $D^{*+}D^{0}$ threshold in 2021 \cite{LHCb:2021vvq}.
Its minimal quark content is $cc\bar{u}\bar{d}$; its width is $\Gamma=48$ keV and its quantum numbers were determined to be $I(J^{P})=0(1^{+})$ \cite{LHCb:2021auc}.

In the sector of fully heavy tetraquark states, 
the LHCb Collaboration reported the observation of two resonances in the $J/\psi J/\psi$ channel in 2020: a broad resonance spanning the range 6.2-6.8 GeV and a narrow resonance centered at approximately 6.9 GeV, with the latter denoted as X(6900) \cite{LHCb:2020bwg}. 
Subsequently, this state was confirmed in the relevant channels by the CMS Collaboration and ATLAS Collaboration.
Beyond X(6900), several other resonance states within the same energy range have been reported, such as X(6600) and X(7200) by the CMS Collaboration \cite{CMS:2023owd,CMS:2025fpt}, as well as X(6200), X(6600), and X(7200) by the ATLAS Collaboration \cite{ATLAS:2023bft}.
All these resonances stand as promising candidates for fully heavy tetraquark states \cite{Wu:2024euj,Huang:2020dci,Lu:2020cns,liu:2020eha,Albuquerque:2020hio,Yang:2020wkh}.

To date, hidden, singly, doubly, and fully heavy tetraquark states have been experimentally observed, while no experimental evidence has been reported for triply heavy tetraquark states.
On the other hand, there have been several theoretical explorations for these types of states using various methods \cite{Li:2025wod,Cheng:2020nho,Yang:2025wqo,Xu:2025zna,Xing:2019wil}.
First, Chen et al. used the chromomagnetic interaction model to systematically investigate the mass gaps of the $QQ\bar{Q}\bar{q}$ tetraquark states and identified several stable states \cite{Chen:2016ont}. 
Moreover, estimations from the chiral quark model for two configurations-i.e., the meson-meson configuration and the diquark-antidiquark configuration-identified several bound triply heavy tetraquark states \cite{Liu:2022jdl}.
Meanwhile, 
lattice QCD calculations indicate that shallowly bound $uc\bar{b}\bar{b}$ and $sc\bar{b}\bar{b}$ states are possible \cite{Hudspith:2020tdf,Junnarkar:2018twb}.
However, estimations from the extended chromomagnetic interaction model \cite{Weng:2021ngd}, the nonrelativistic quark model \cite{Silvestre-Brac:1993zem}, and the extended relativized quark model \cite{Lu:2021kut} indicate that there is no bound triply heavy tetraquark state.
Yang et al. predicted narrow resonant states in all allowed $I(J^{P})$ quantum channels for the $S$-wave triply heavy tetraquark states $cc\bar{c}\bar{q}$ and $bb\bar{b}\bar{q}$ within the constituent quark model \cite{Yang:2024nyc}.
Furthermore, QCD sum rule analyzes indicate that $bb\bar{b}\bar{q}$ states are expected to be stable against strong decays \cite{Jiang:2017tdc}.
Additional theoretical calculations have also estimated the mass spectra, magnetic moments,
charge radii, and decay properties of $QQ\bar{Q}\bar{q}$ tetraquark states \cite{Zhang:2024jvv,Galkin:2025ubt,Li:2025fmf,Yang:2025jsp,Meng:2023jqk,Zhu:2023lbx,Mutuk:2023yev}.
In summary, the existence of triply heavy tetraquark states remains an open question.

Given the absence of experimental evidence and the aforementioned theoretical discrepancies, this study focuses on four types of triply heavy tetraquark states, namely $cc\bar{c}\bar{n}$, $cc\bar{c}\bar{s}$, $bb\bar{b}\bar{n}$, and $bb\bar{b}\bar{s}$ ($n=u,d$), to systematically investigate their mass spectra and decay properties. 
Based on the nonrelativistic quark model, we precisely solve the four-body Schr\"{o}dinger equation using the Gaussian expansion method, perform the expansion of the spatial part of the wave function with Jacobi coordinates, and consider the effect of color-spin configuration mixing. 
Meanwhile, the quark-interchange model is employed to calculate the decay widths of OZI-allowed two-body rearrangement strong decays, and the tree-level quark-photon coupling is used to derive the radiative decay widths, 
thereby comprehensively revealing the key properties of such triply heavy tetraquark states.

The paper is organized as follows. 
After the Introduction, Section \ref{sec2} elaborates on the effective Hamiltonian and the total wave functions (spatial, flavor, color, and spin parts).
Section \ref{sec3} presents the Gaussian expansion method for solving the four-body Schr\"{o}dinger equation, the definition and calculation of root-mean-square (RMS) radii, and the theoretical frameworks for radiative decays and OZI-allowed two-body rearrangement strong decays. 
Section \ref{sec4} provides detailed numerical results and in-depth discussions, focusing on the mass spectra, internal spatial structures, and decay properties of the $cc\bar{c}\bar{n}$, $cc\bar{c}\bar{s}$, $bb\bar{b}\bar{n}$, and $bb\bar{b}\bar{s}$ systems. 
Finally, Section \ref{sec5} summarizes the key findings of this work.

\section{Hamiltonian and wave functions}\label{sec2}

\subsection{The effective Hamiltonian
}\label{sec21}
In the nonrelativistic quark model, 
the Hamiltonian for the ground-state tetraquark system includes confinement potential and hyperfine potential for color-spin interactions \cite{Park:2016mez,Park:2017jbn,Noh:2021lqs,Park:2023ygm,Park:2024gbq,An:2022qpt,An:2022fvs,An:2025qfw,An:2025rjv}:
\begin{eqnarray}\label{Eq1}
H=\sum_{i=1}^{4}(m_{i}+\frac{\textbf{p}^{2}_{i}}{2m_{i}})-\frac{3}{4}\sum_{i<j}^{4}
\frac{\lambda^{c}_{i}}{2}.\frac{\lambda^{c}_{j}}{2}(V^{\rm C}_{ij}+V^{\rm CS}_{ij}),
\end{eqnarray}
where $m_{i}$ is the quark mass, $\textbf{p}^{2}_{i}/(2m_{i})$ represents the kinetic energy,
and $\lambda^{c}_{i}/2$ is the SU(3) color operator for the $i$-th quark.
For an antiquark, $\lambda^{c}_{i}$ should be replaced by $-\lambda^{c*}_{i}$. 
Here, the confinement potential $V^{{\rm C}}_{ij}$ and hyperfine potential 
$V^{{\rm CS}}_{ij}$ are defined as:
\begin{eqnarray}\label{Eq2}
V^{{\rm C}}_{ij}&=&-\frac{\kappa}{r_{ij}}+\frac{r_{ij}}{a^{2}_{0}}-C,\nonumber\\
V^{{\rm CS}}_{ij}&=&\frac{\kappa'}{m_{i}m_{j}}\frac{1}{r_{0ij}r_{ij}}e^{-r^{2}_{ij}/r^{2}_{0ij}}\sigma_{i}\cdot\sigma_{j},
\end{eqnarray}
where $r_{ij}=|\textbf{r}_{i}-\textbf{r}_{j}|$ is the relative distance between the $i$-th and the $j$-th 
(anti)quarks, and $\sigma_{i}$ stands for the SU(2) spin operator for the $i$-th (anti)quark.
The parameter $C$ represents a mass-renormalized constant.
In the heavy quark mass limit ($m_{c,b}\to +\infty$), $V^{{\rm CS}}_{ij}$ reduces to  $1/(m_{i}m_{j}) \delta({r_{ij}})$; this form is chosen to fit the mass gaps of conventional hadrons containing both light and heavy quarks.
Furthermore, the $r_{0ij}$ and $\kappa'$ are taken to depend on the quark masses, given by
\begin{eqnarray}\label{Eq3}
r_{0ij}&=&1/(\alpha+\beta\frac{m_{i} m_{j}}{m_{i}+m_{j}}),\nonumber\\
\kappa'&=&\kappa_{0}(1+\gamma\frac{m_{i} m_{j}}{m_{i}+m_{j}}).
\end{eqnarray}

The numerical values of the parameters in Eqs.(\ref{Eq1}-\ref{Eq3}) are determined by fitting to the experimental masses of single-charmed (bottom) mesons and charmonium (bottomonium) using the Gaussian expansion method.
These parameters are listed in Table \ref{para}.
For completeness, Table \ref{para} also presents the theoretical and experimental masses of charmed (bottom) mesons and charmonium (bottomonium), along with their corresponding errors and root-mean-square (RMS) radii, for comparative analysis.
\\

\begin{table*}[t]
\caption{
Parameters of the Hamiltonian determined by fitting the singly-charmed (bottom) mesons and charmonium (bottomonium) masses.
The $M_{\rm the}$, $M_{\rm exp}$, Error, and RMS Radius denote the theoretical value, the experimental value, the discrepancy, and the root-mean-square radii, respectively.
}\label{para}
\begin{lrbox}{\tablebox}
\renewcommand\arraystretch{1.85}
\renewcommand\tabcolsep{2.8pt}
\begin{tabular}{c|cccccccccc}
\toprule[1.50pt]
\toprule[0.50pt]
Parameter&\multicolumn{1}{c|}{$m_{n}$}&\multicolumn{1}{c|}{$m_{s}$}&\multicolumn{2}{c|}{$a_{0}$}&\multicolumn{2}{c|}{$\beta$}&
\multicolumn{2}{c}{$\kappa_{0}$}\\
\Xcline{1-9}{0.3pt}
Value&\multicolumn{1}{c|}{466.0 MeV}&\multicolumn{1}{c|}{681.0 MeV}&\multicolumn{2}{c|}{$3.26\times10^{-2}$ $\rm (MeV^{-1}fm)^{1/2}$}& \multicolumn{2}{c|}{$1.31\times10^{-4}$ $\rm (MeV fm)^{-1}$}&\multicolumn{2}{c}{$2.12\times10^{2}$ MeV}\\
\Xcline{1-9}{0.3pt}
Parameter&\multicolumn{1}{c|}{$m_{c}$}&\multicolumn{1}{c|}{$m_{b}$}&\multicolumn{1}{c|}{$C$}&\multicolumn{1}{c|}{$\alpha$}&\multicolumn{2}{c|}{$\kappa$}&\multicolumn{2}{c}{$\gamma$}\\
\Xcline{1-9}{0.3pt}
Value&\multicolumn{1}{c|}{1933.0 MeV}&\multicolumn{1}{c|}{5345.0 MeV}& \multicolumn{1}{c|}{1013.2 MeV}&\multicolumn{1}{c|}{1.09 $\rm fm^{-1}$}&\multicolumn{2}{c|}{$1.08\times10^{2}$ MeV fm}&\multicolumn{2}{c}{$1.26\times10^{-3}$ $\rm MeV^{-1}$}\\
\midrule[1.5pt]
Meson&\multicolumn{1}{c|}{$D$}&\multicolumn{1}{c|}{$D^{*}$}&\multicolumn{1}{c|}{$D_{s}$}&\multicolumn{1}{c|}{$D^{*}_{s}$}&\multicolumn{1}{c|}{$\eta_{c}$}&\multicolumn{1}{c|}{$J/\psi$}\\
\Xcline{1-7}{0.3pt}
$M_{\rm the}$ (MeV)&\multicolumn{1}{c|}{1856.9}&\multicolumn{1}{c|}{1994.4}&\multicolumn{1}{c|}{1969.3}&\multicolumn{1}{c|}{2095.9}&\multicolumn{1}{c|}{2999.8}&\multicolumn{1}{c|}{3095.8}\\
\Xcline{1-7}{0.3pt}
$M_{\rm exp}$ (MeV)&\multicolumn{1}{c|}{1869.7}&\multicolumn{1}{c|}{2010.3}&\multicolumn{1}{c|}{1968.3}&\multicolumn{1}{c|}{2112.2}&\multicolumn{1}{c|}{\quad2983.6\quad\quad}&\multicolumn{1}{c|}{3096.9}\\
\Xcline{1-7}{0.3pt}
Error (MeV)&\multicolumn{1}{c|}{-12.8}&\multicolumn{1}{c|}{-15.9}&\multicolumn{1}{c|}{1.0}&\multicolumn{1}{c|}{-16.3}&\multicolumn{1}{c|}{16.2\quad}&\multicolumn{1}{c|}{-1.1}\\
\Xcline{1-7}{0.3pt}
RMS Radius (fm)&\multicolumn{1}{c|}{0.47}&\multicolumn{1}{c|}{0.46}&\multicolumn{1}{c|}{0.42}&\multicolumn{1}{c|}{0.46}&\multicolumn{1}{c|}{0.31\quad}&\multicolumn{1}{c|}{0.33}\\
\Xcline{1-7}{1pt}
Meson&\multicolumn{1}{c|}{$B$}&\multicolumn{1}{c|}{$B^{*}$}&\multicolumn{1}{c|}{$B_{s}$}&\multicolumn{1}{c|}{$B^{*}_{s}$}&\multicolumn{1}{c|}{$\eta_{b}$}&\multicolumn{1}{c|}{$\Upsilon$}\\
\Xcline{1-7}{0.3pt}
$M_{\rm the}$ (MeV)&\multicolumn{1}{c|}{5281.8}&\multicolumn{1}{c|}{5337.2}&\multicolumn{1}{c|}{5368.6}&\multicolumn{1}{c|}{5423.3}&\multicolumn{1}{c|}{9397.8}&\multicolumn{1}{c|}{9459.3}\\
\Xcline{1-7}{0.3pt}
$M_{\rm exp}$ (MeV)&\multicolumn{1}{c|}{5279.3}&\multicolumn{1}{c|}{5325.2}&\multicolumn{1}{c|}{5366.8}&\multicolumn{1}{c|}{5415.4}
&\multicolumn{1}{c|}{9398.0}&\multicolumn{1}{c|}{9460.3}\\
\Xcline{1-7}{0.3pt}
Error (MeV)&\multicolumn{1}{c|}{2.6}&\multicolumn{1}{c|}{12.0}&\multicolumn{1}{c|}{1.8}&\multicolumn{1}{c|}{7.9}&\multicolumn{1}{c|}{-0.2}&\multicolumn{1}{c|}{-1.0}\\
\Xcline{1-7}{0.3pt}
RMS Radius (fm)&\multicolumn{1}{c|}{0.47}&\multicolumn{1}{c|}{0.49}&\multicolumn{1}{c|}{0.40}&\multicolumn{1}{c|}{0.42}&\multicolumn{1}{c|}{0.16\quad}&\multicolumn{1}{c|}{0.17}\\
\Xcline{1-7}{1pt}
\bottomrule[0.50pt]
\bottomrule[1.50pt]
\end{tabular}
\end{lrbox}\scalebox{1.00}{\usebox{\tablebox}}
\end{table*}

\subsection{The wave functions}\label{sec22}

The total wave function consists of four parts: the spatial part, flavor part, color part, and spin part.
\begin{eqnarray}\label{total}
\Psi_{tot}=\Psi_{spatial}\otimes F_{flavor}\otimes \psi_{color}\otimes \chi_{spin}.
\end{eqnarray}
In the flavor part, the investigated triply-charmed (triply-bottom) tetraquark systems correspond to four distinct flavor configurations: $cc\bar{c}\bar{n}$, $cc\bar{c}\bar{s}$, $bb\bar{b}\bar{n}$, and $bb\bar{b}\bar{s}$ ($n=u,d$).
Since the first two quarks in the above configurations are identical, 
we adopt a diquark-antidiquark picture to analyze these tetraquark systems, 
which facilitates the discussion of the constraints imposed by the Pauli principle.

In the color part, the color wave functions are analyzed using SU(3) group theory.
Meanwhile, due to the requirement of color confinement, the color wave function must be a color singlet.
In the diquark-antidiquark picture, the color decomposition reads:
\begin{eqnarray}\label{color}
&&(3\otimes3)\otimes(\bar{3}\otimes\bar{3})\nonumber\\
&=&(\bar{3}\oplus6)\otimes(3\oplus\bar{6})\to(\bar{3}\otimes3)\oplus(6\otimes\bar{6}).
\end{eqnarray}
Based on this, we obtain two types of color-singlet states:
\begin{eqnarray}\label{color1}
\psi_{1}=|(QQ)^{\bar{3}_{c}}(\bar{Q}\bar{q})^{3_{c}}\rangle,
\quad
\psi_{2}=|(QQ)^{6_{c}}(\bar{Q}\bar{q})^{\bar{6}_{c}}\rangle.
\end{eqnarray}
In the notation $|(QQ)^{\rm color_{1}}(\bar{Q}\bar{q})^{\rm color_{2}}\rangle$,
the $\rm color_{1}$ and $\rm color_{2}$ denote the color representations of the diquark $(QQ)$ and the antidiquark $(\bar{Q}\bar{q})$, respectively.

The spin part is based on SU(2) symmetry;
all possible wave functions in the diquark-antidiquark picture are as follows:
\begin{eqnarray}\label{spin1}
\chi_{1}&=&|(QQ)_{1}(\bar{Q}\bar{q})_{1}\rangle_{2},
\quad
\chi_{2}=|(QQ)_{1}(\bar{Q}\bar{q})_{1}\rangle_{1},\nonumber\\
\chi_{3}&=&|(QQ)_{1}(\bar{Q}\bar{q})_{0}\rangle_{1},
\quad
\chi_{4}=|(QQ)_{0}(\bar{Q}\bar{q})_{1}\rangle_{1},\nonumber\\
\chi_{5}&=&|(QQ)_{1}(\bar{Q}\bar{q})_{1}\rangle_{0},
\quad
\chi_{6}=|(QQ)_{0}(\bar{Q}\bar{q})_{0}\rangle_{0}.
\end{eqnarray}
In the notation $|(QQ)_{\rm spin_{1}}(\bar{Q}\bar{q})_{\rm spin_{2}}\rangle_{\rm spin_{3}}$, $\rm spin_{1}$, $\rm spin_{2}$, and $\rm spin_{3}$ correspond to the spin of the diquark $(QQ)$, the spin of the antidiquark $(\bar{Q}\bar{q})$, and the total spin of the tetraquark state, respectively.

\begin{figure}[t]
\includegraphics[width=0.98\linewidth]{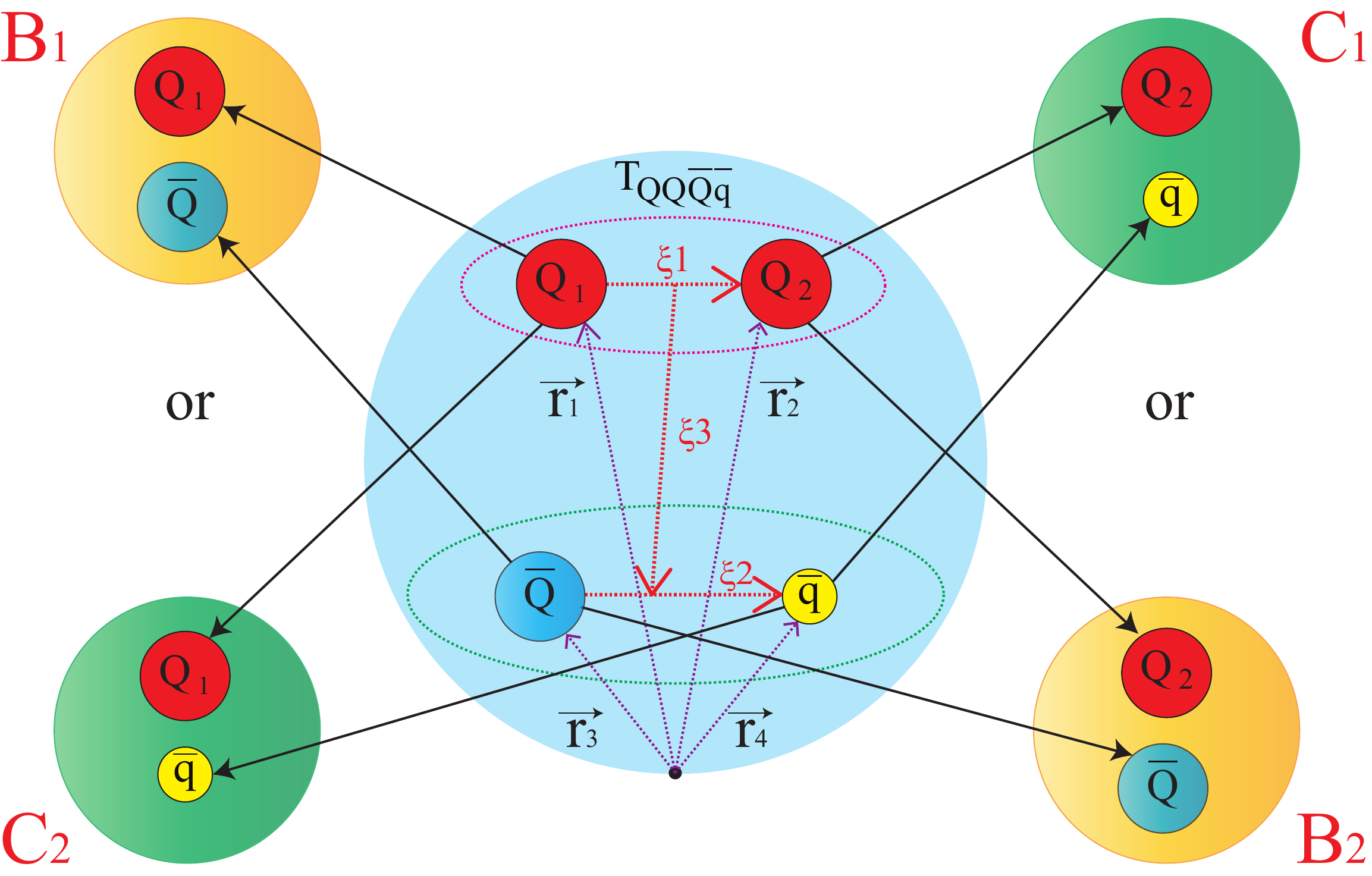}
\caption{
Spatial coordinates defined for the triply heavy tetraquark system $QQ\bar{Q}\bar{q}$ and its two-body rearrangement decays into meson-meson final states via quark interchange. Here, the meson-meson final state can form via two quark rearrangement pathways $B_{1}C_{1}$ ($(Q_{1}\bar{Q})(Q_{2}\bar{q})$) and $B_{2}C_{2}$ ($(Q_{2}\bar{Q})(Q_{1}\bar{q})$), as illustrated in the figure.
}\label{fig2}
\end{figure}

In the spatial part, since we only consider low-lying $S$-wave $QQ\bar{Q}\bar{q}$ tetraquark states, 
the spatial wave function is symmetric under the exchange of two identical heavy quarks.
We employ the Gaussian expansion method (GEM) to accurately solve the four-body problem \cite{Hiyama:2003cu,Hiyama:2012sma}.
The relative Jacobi coordinates for the four-body system are defined in terms of the single-particle coordinates $\vec{r}_{i}$ $(i=1,2,3,4)$ as follows (see Fig. \ref{fig2}):
\begin{eqnarray}\label{jacobi}
&&\xi_{1}=\sqrt{1/2}(\vec{r}_{1}-\vec{r}_{2}),\nonumber\\
&&\xi_{2}=\sqrt{1/2}(\vec{r}_{3}-\vec{r}_{4}),\nonumber\\
&&\xi_{3}=(\frac{m_{1}\vec{r}_{1}+m_{2}\vec{r}_{2}}{m_{1}+m_{2}})-(\frac{m_{3}\vec{r}_{3}+m_{4}\vec{r}_{4}}{m_{3}+m_{4}}),\nonumber\\
&&\textbf{R}=\frac{m_{1}\vec{r}_{1}+m_{2}\vec{r}_{2}+m_{3}\vec{r}_{3}+m_{4}\vec{r}_{4}}{m_{1}+m_{2}+m_{3}+m_{4}}.
\end{eqnarray}
Here, $\xi_{1}$ ($\xi_{2}$) corresponds to the relative Jacobi coordinate between the two heavy quarks $Q$ (between the heavy antiquark $\bar{Q}$ and the light antiquark $\bar{q}$).
Meanwhile, $\xi_{3}$ corresponds to the relative Jacobi coordinate 
between the centers of mass of the diquark $(QQ)$ and the antidiquark $(\bar{Q}\bar{q})$.

Here, introducing the center-of-mass frame ($\textbf{R}=0$) is quite beneficial. 
The number of independent Jacobi coordinates is reduced to three, which allows  us to appropriately reduce the kinetic term of the Hamiltonian in Eq. (\ref{Eq1}) and facilitates our calculations.
The kinetic term in the center-of-mass frame, denoted as $T_{c}$, takes the explicit form:
\begin{eqnarray}\label{kinetic term}
T_{c}=\sum^{4}_{i=1}\frac{\textbf{p}^{2}_{\vec{r}_{i}}}{2m_{i}}- \frac{\textbf{p}^{2}_{R}}{2M}= \frac{\textbf{p}^{2}_{\xi_{1}}}{2m'_{1}}+\frac{\textbf{p}^{2}_{\xi_{2}}}{2m'_{2}}+\frac{\textbf{p}^{2}_{\xi_{3}}}{2m'_{3}},
\end{eqnarray}
where the reduced masses $m'_{i}$ $(i=1,2,3)$ are defined as:
\begin{eqnarray}\label{kinetic term2}
&&m'_{1}=\frac{2\times (m_{1}m_{2})}{m_{1}+m_{2}}, \quad
m'_{2}=\frac{2\times (m_{3}m_{4})}{m_{3}+m_{4}}, \nonumber\\
&&m'_{3}=\frac{ (m_{1}+m_{2}) \times (m_{3}+m_{4})}{m_{1}+m_{2}+m_{3}+m_{4}}.
\end{eqnarray}

Based on the Jacobi coordinates in the center-of-mass frame ($\textbf{R}=0$) defined in Eq.(\ref{jacobi}), 
the spatial part is expanded in a set of correlated Gaussian basis functions:
\begin{eqnarray}\label{spatial}
&&\Psi_{spatial}(\xi_{1},\xi_{2},\xi_{3})\nonumber\\
&=&\sum^{n_{1max}}_{n_{1}=1}\sum^{n_{2max}}_{n_{2}=1}\sum^{n_{3max}}_{n_{3}=1}
c_{n_{1}n_{2}n_{3}}\psi^{n_{1}n_{2}n_{3}}(\xi_{1},\xi_{2},\xi_{3})\nonumber\\
&=&\sum^{n_{1max}}_{n_{1}=1}\sum^{n_{2max}}_{n_{2}=1}\sum^{n_{3max}}_{n_{3}=1}
c_{n_{1}n_{2}n_{3}}\rm Exp[-a_{n_{1}}\xi^{2}_{1}-a_{n_{2}}\xi^{2}_{2}-a_{n_{3}}\xi^{2}_{3}],\nonumber\\
\end{eqnarray}
where the expansion coefficients $c_{n_{1}n_{2}n_{3}}$ are determined via the Rayleigh-Ritz variational method.
The $\xi_{i}$ $(i=1,2,3)$ are Jacobi coordinates, and the Gaussian range parameters $a_{n_{i}}$ $(i=1,2,3)$ are defined as:
\begin{eqnarray}\label{spatial4}
a_{n_{i}}=1/r^{2}_{n_{i}}, r_{n_{i}}=r_{min_{i}}d^{n_{i}-1} (n_{i}=1,2,...,n_{max_{i}}), 
\end{eqnarray}
where the ratio coefficient $d$ is given by
\begin{eqnarray}\label{spatial41}
d=(\frac{r_{max_{i}}}{r_{min_{i}}})^{\frac{1}{n_{max_{i}}-1}} \quad (i=1,2,3).
\end{eqnarray}
There are three parameters \{$r_{max_{i}}$, $r_{min_{i}}$, $n_{max_{i}}$\} to be determined through the variational method to minimize the total energy.
Stable numerical results can be obtained by choosing the parameters \{5 fm, 0.7 fm, 5\}.
These numerical results are independent of the parameters \{$r_{max_{i}}$, $r_{min_{i}}$, $n_{max_{i}}$\}.
To verify this parameter independence, we further adjust the value of $n_{max_{i}}$ to 6 and 7, confirming that the calculated masses are nearly identical.

Combining spatial part and flavor combinations, we present all possible color-spin configurations with different $J^{P}$ quantum numbers that satisfy antisymmetry, as listed in Table \ref{QQQq}.
In Table \ref{QQQq}, the notation $|(QQ)^{\rm color_{1}}_{\rm spin_{1}}(\bar{Q}\bar{q})^{\rm color_{2}}_{\rm spin_{2}}\rangle_{\rm spin_{3}}$ is adopted to denote the total wave function.

\section{Theoretical Framework: Mass Spectra and Decay Properties}\label{sec3}
\subsection{Gaussian expansion method and Root-mean-square radii}\label{secaa}
Using the preparations outlined in Sec.\ref{sec2}, 
the mass spectra are obtained by solving the four-body Schr\"{o}dinger equation via the Gaussian expansion method (GEM):
\begin{eqnarray}\label{equa}
\hat{H}\Psi_{\rm total}(\xi_{1},\xi_{2},\xi_{3})=E\Psi_{\rm total}(\xi_{1},\xi_{2},\xi_{3}),
\end{eqnarray}
where $\Psi_{\rm total}(\xi_{1},\xi_{2},\xi_{3})$ corresponds to the total wave function defined in Eq.(\ref{total}), $\hat{H}$ represents the Hamiltonian operator given in Eq.(\ref{Eq1}), and $E$ denotes the energy eigenvalues associated with the tetraquark mass spectra.

The kinetic, potential, and overlap matrix elements are defined as follows:
\begin{eqnarray}\label{acca}
&&T_{c}^{nn'}=\langle\Psi^{n_{1}n_{2}n_{3}}(\xi_{1},\xi_{2},\xi_{3})|T_{c}|\Psi^{n'_{1}n'_{2}n'_{3}}(\xi_{1},\xi_{2},\xi_{3})\rangle,\nonumber\\
&&V^{nn'}_\alpha=
\langle\Psi^{n_{1}n_{2}n_{3}}(\xi_{1},\xi_{2},\xi_{3})\psi_{cs}|
V_\alpha|\psi'_{cs}\Psi^{n'_{1}n'_{2}n'_{3}}(\xi_{1},\xi_{2},\xi_{3})\rangle ,\nonumber\\
&&N^{nn'}=\langle\Psi^{n_{1}n_{2}n_{3}}(\xi_{1},\xi_{2},\xi_{3})\psi_{cs}|\psi'_{cs}\Psi^{n'_{1}n'_{2}n'_{3}}(\xi_{1},\xi_{2},\xi_{3})\rangle.\nonumber\\
\end{eqnarray}
Here, $\Psi^{n_{1}n_{2}n_{3}}(\xi_{1},\xi_{2},\xi_{3})$ denotes the spatial part given in Eq.(\ref{spatial}), and $\psi_{cs}$ denotes the color-spin wave function, which is listed in Table \ref{QQQq}.
Meanwhile, $T_{c}$ is kinetic term given in Eq.(\ref{kinetic term}), and $V_\alpha$ corresponds to $V^{\rm C}$ and $V^{\rm CS}$ given in Eq.(\ref{Eq2}), where $\alpha=1,...,6$ labels the number of pairwise inter(anti)quark interaction potentials.
According to Eq.~(\ref{acca}), the Schr\"{o}dinger equation Eq.~(\ref{equa}) can be transformed into a generalized matrix eigenvalue problem by applying the Rayleigh–Ritz variational principle, which reads
\begin{eqnarray} \label{tven}
[T^{nn'}_{c}+\sum^{6}_{\alpha=1}V^{nn'}_{\alpha}]C_{nn'}=EN^{nn'}C_{nn'}.
\end{eqnarray}
Solving this eigenvalue problem yields the tetraquark mass spectra and their corresponding internal energy contributions, as summarized in Table \ref{QQQq}.

The RMS radius $\langle r^{2}_{ij} \rangle$, which denotes the average distance between two (anti)quarks, is defined as follows \cite{Liu:2024fnh}:
\begin{eqnarray}\label{rms}
\langle r^{2}_{ij} \rangle=\int (\vec{r}_{i}-\vec{r}_{j})^{2}|\Psi_{spatial}(\xi_{1},\xi_{2},\xi_{3})|^{2}d\xi_{1}d\xi_{2}d\xi_{3}.
\end{eqnarray}
The calculated RMS radii for various flavor configurations are tabulated in Table \ref{QQQq3}.
For further comparison, the RMS radii of conventional mesons are also presented in Table \ref{para}.
In Table \ref{QQQq3}, 
$\langle r^{2}_{12} \rangle^{\frac{1}{2}}$ quantifies the RMS spatial separation between two heavy quarks ($QQ$).
Similarly,
$\langle r^{2}_{34} \rangle^{\frac{1}{2}}$ quantifies the RMS spatial separation between light antiquark $\bar{q}$ and heavy antiquark $\bar{Q}$.
$\langle r^{2}_{13} \rangle^{\frac{1}{2}}$ and $\langle r^{2}_{23} \rangle^{\frac{1}{2}}$ describe the RMS spatial separation between a heavy quark $Q$ and light antiquark $\bar{q}$,
while 
$\langle r^{2}_{14} \rangle^{\frac{1}{2}}$ and $\langle r^{2}_{24} \rangle^{\frac{1}{2}}$ describe the RMS spatial separation between a heavy quark $Q$ and heavy antiquark $\bar{Q}$.
Furthermore, 
$\langle r^{2}_{12-34} \rangle^{\frac{1}{2}}$ denotes the average distance between the mass centers of the heavy diquark ($QQ$) and the antidiquark ($\bar{Q}\bar{q}$), while 
$\langle r^{2}_{13-24} \rangle^{\frac{1}{2}}$ and $\langle r^{2}_{14-23} \rangle^{\frac{1}{2}}$ describe the RMS spatial separation between two quark-antiquark pairs ($Q\bar{Q}-Q\bar{q}$).

Owing to the identical particles of the two heavy quarks in this study, symmetry dictates the following relations:
$\langle r^{2}_{13} \rangle^{\frac{1}{2}}=\langle r^{2}_{23} \rangle^{\frac{1}{2}}$, $\langle r^{2}_{14} \rangle^{\frac{1}{2}}=\langle r^{2}_{24} \rangle^{\frac{1}{2}}$, and $\langle r^{2}_{13-24} \rangle^{\frac{1}{2}}=\langle r^{2}_{14-23} \rangle^{\frac{1}{2}}$.
Notably, if the average separation between the two subclusters is on the same scale as or even smaller than that between any two constituent (anti)quarks ($\langle r^{2}_{ij-kl}\rangle^{1/2}\sim\langle r^{2}_{ij}\rangle^{1/2}$), it indicates a high degree of spatial overlap between these substructures \cite{Liu:2024fnh}.
Such pronounced overlap supports a compact tetraquark interpretation for the state.
Leveraging this key characteristic, we can probe the spatial nature of the tetraquark system by comparing RMS separations.

\subsection{Radiative decay}\label{secbb}

Besides the mass spectra and RMS radii, we further calculate the decay properties, including radiative decays and rearrangement strong decays, using the obtained wave functions.
Refs. \cite{Davila-Rivera:2025exk,Wang:2023ael,Peng:2024pyl,Zhou:2025fpp,Sheng:2024hkf,Zhang:2025ame} have well established the theoretical framework for calculating radiative decays within the nonrelativistic constituent quark model, 
and this formalism has been widely and successfully applied to various hadronic systems.

The coupling between quarks and photons at the tree level can be written as:
\begin{eqnarray}\label{rad}
H_e = -\sum_j e_j \bar{\psi}_j \gamma_\mu^j A^\mu(\mathbf{k}, \vec{r}_j) \psi_j,
\end{eqnarray}
where $\psi_{j}$ and $A^\mu(\mathbf{k}, \vec{r}_j)$ correspond to the $j$-th quark field and photon field, respectively.
In addition, $\vec{r}_j$, $e_{j}$, and $\gamma_\mu^j$ are the coordinate, electric charge, and Dirac matrices of the 
$j$-th quark, and $\mathbf{k}$ is the momentum of the photon.

Through the nonrelativistic approximation, we obtain \cite{Li:1994cy,Li:1997gd,Zhao:2002id}:
\begin{eqnarray}\label{rad1}
H_e^{\text{nr}} = \sum_j \left[ e_j \vec{r}_j \cdot \boldsymbol{\epsilon} - \frac{e_j}{2m_j} \boldsymbol{\sigma}_j \cdot (\boldsymbol{\epsilon} \times \hat{\mathbf{k}}) \right] e^{-i\mathbf{k} \cdot \vec{r}_j},
\end{eqnarray}
where $m_{j}$ and $\boldsymbol{\sigma}_{j}$ are the mass and Pauli matrix of the $j$-th quark, respectively, and
$\boldsymbol{\epsilon}$ is the polarization vector of the photon. 
Based on this, the amplitude for the hadron radiative decay process is calculated as \cite{Deng:2016stx,Deng:2016ktl}:
\begin{eqnarray}\label{rad2}
\mathcal{M} = -i \sqrt{\frac{\omega_\gamma}{2}} \langle \Psi_{H_f} | H_e^{\text{nr}} | \Psi_{H_i} \rangle,
\end{eqnarray}
where $|\Psi_{H_{i}}\rangle$ and $|\Psi_{H_{f}}\rangle$ are the total wave functions of the initial and final hadrons, respectively, and $\omega_\gamma$ is the energy of the photon, which is defined as follows:
\begin{eqnarray}\label{rad3}
\omega_\gamma=\frac{M^{2}_{i}-M^{2}_{f}}{2M_{i}},
\end{eqnarray}
where $M_{i}$ and $M_{f}$ are the masses of the initial and final hadron, respectively.

Here, we choose the coordinate system such that the photon momentum aligns with the $z$-axis [$\mathbf{k}=(0,0,\omega_\gamma)$] and take the polarization vector as $\boldsymbol{\epsilon}=-1/\sqrt{2}(1,i,0)$.
The radiative decay width is then obtained from the helicity
amplitude via \cite{Xiao:2017udy,Lu:2017meb,Wang:2017kfr,Yao:2018jmc}:
\begin{eqnarray}\label{rad4}
\Gamma = \frac{|\mathbf{k}|^2}{\pi} \frac{2}{2J_i + 1} \frac{M_f}{M_i} \sum_{M_{J_f}, M_{J_i}} \left| \mathcal{M}_{M_{J_f}, M_{J_i}} \right|^2,
\end{eqnarray}
where $J_i$ and $J_f$ are the total angular momenta of the initial and final hadrons, respectively.
Finally, we calculate the corresponding radiative decay widths via Eq.~(\ref{rad4}) and tabulate them in Table \ref{QQQq3}.

\subsection{Rearrangement decay}\label{seccc}

\begin{figure*}[htbp]
\includegraphics[width=0.98\linewidth]{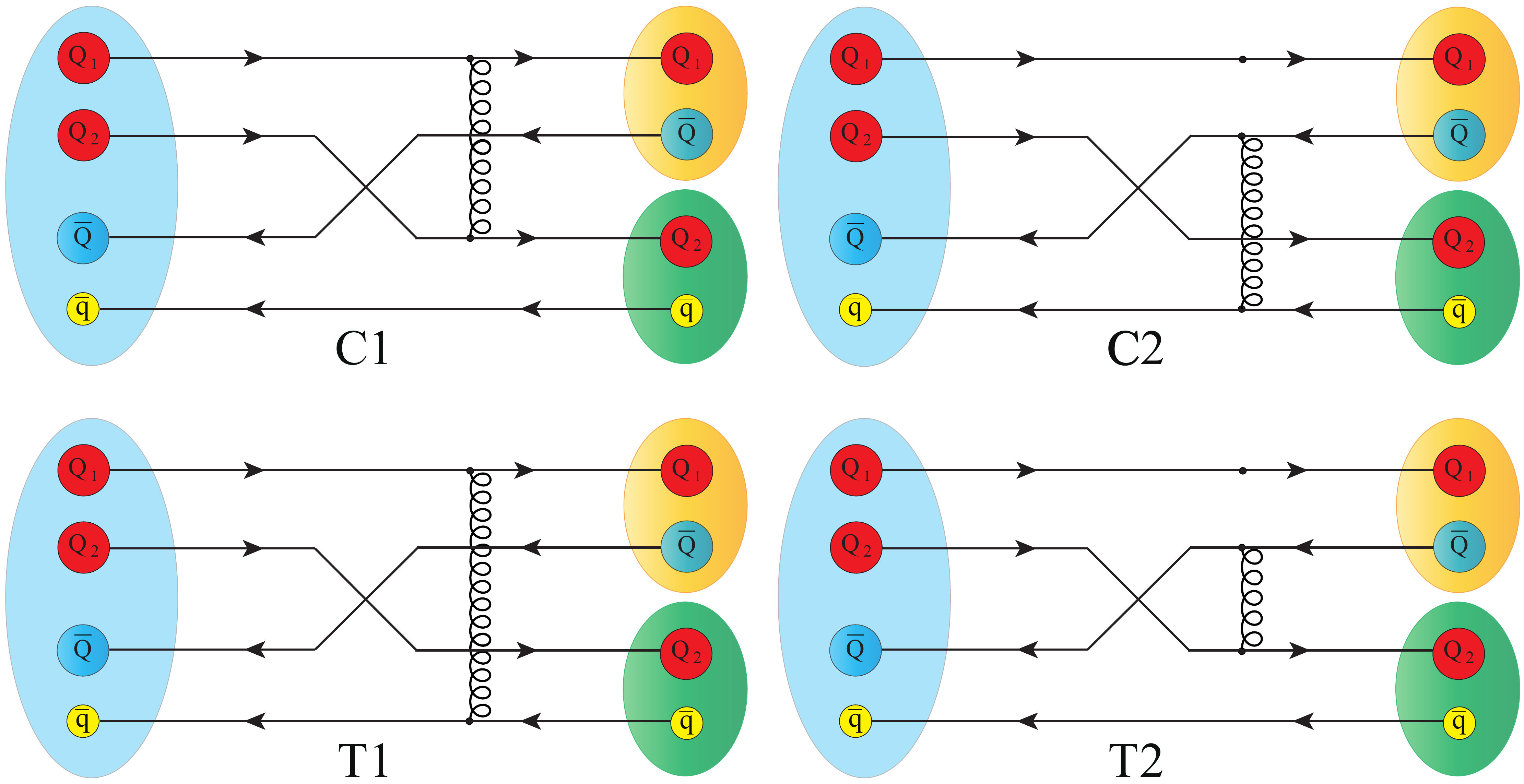}
\caption{
Quark-interchange diagrams for the triply heavy tetraquark $QQ\bar{Q}\bar{q}$ decaying into meson-meson final states at the quark level. 
The curly line denotes the quark-(anti)quark interactions.
}\label{fig3}
\end{figure*}

In addition to radiative decays,
we also calculate the OZI-allowed two-body strong decays of the $cc\bar{c}\bar{q}$ and $bb\bar{b}\bar{q}$ tetraquark states within the quark-interchange model \cite{Barnes:1991em,Wong:2001td}.
This well-established model has been successfully employed to describe the rearrangement decays of a wide range of exotic hadronic states \cite{Wang:2020prk,Zhou:2019swr,Liu:2022hbk,Xiao:2019spy,Wang:2019spc,Liang:2024met,Wang:2018pwi,Yang:2021sue,Barnes:2000hu}.
Where phase space allows, the primary two-body strong decay for these tetraquark states is the rearrangement process:
$QQ\bar{Q}\bar{q}\to Q\bar{Q}+Q\bar{q}$, as illustrated in Fig. \ref{fig2}.

The partial decay width $\Gamma$ for the decay process 
$A\to BC$ is given by
\begin{eqnarray}\label{width}
\Gamma=\frac{1}{(2J_{A}+1)}\frac{|\vec{P}_{B}|}{32\pi^{2}M^{2}_{A}}\int d\Omega|\mathcal{M}(A \to BC)|^{2},
\end{eqnarray}
where $A$ denotes the initial triply heavy tetraquark state, 
while $B$ and $C$ correspond to the final color-singlet mesons: the heavy quarkonia $Q\bar{Q}$ and singly heavy meson $Q\bar{q}$, respectively (see Fig. \ref{fig2}).
Here, $\vec{P}_{B}$ is the three-vector momentum of particle $B$ or $C$ in the initial-hadron-rest frame, and $M_{A}$ is the mass of the initial triply heavy tetraquark state.
The decay amplitude $\mathcal{M}(A \to BC)$ for the tetraquark state is given by
\begin{eqnarray}\label{amp}
\mathcal{M}(A \to BC)=
-(2\pi)^{3/2}\sqrt{2M_{A}}\sqrt{2E_{B}}\sqrt{2E_{C}}\times T,
\end{eqnarray}
where $E_{B}$ and $E_{C}$ are the energies of the final states $B$ and $C$ in the initial-hadron-rest frame, respectively.
The $T$-matrix is expressed as
\begin{eqnarray}\label{Eq:T1}
T&=&\langle\Psi^{B}\Psi^{C}|\sum_{i<j}V_{ij}|\Psi^{A}\rangle\nonumber\\
&=&\langle\Psi^{B}\Psi^{C}|\sum_{i<j}V_{ij}|\Psi^{A}_{(QQ)}\Psi^{A}_{(\bar{Q}\bar{q})}\Psi^{A}_{(QQ)-(\bar{Q}\bar{q})}\rangle,\quad
\end{eqnarray}
where $\Psi^{A}$, $\Psi^{B}$, and $\Psi^{C}$ represent the total wave functions of the initial triply heavy tetraquark state (see Subsec. \ref{sec22}), final heavy quarkonia $Q\bar{Q}$, and final singly heavy meson $Q\bar{q}$, respectively.
Here, $V_{ij}$ represents the potential given in Eq.~(\ref{Eq2}).

Furthermore, the $T$-matrix in momentum space can be rewritten as
\begin{align}\label{Eq:T2}
T = \frac{1}{(2\pi)^{3}}\int d\vec{P}_{\alpha}V_{\rm eff}(\vec{P}_{\alpha},\vec{P}_{B})\Psi^{A}_{(QQ)-(\bar{Q}\bar{q})}(\vec{P}_{\alpha}),
\end{align}
where the effective potential $V_{\rm eff}(\vec{P}_{\alpha},\vec{P}_{B})$ is a function of the initial and final relative momenta $\vec{P}_{\alpha}$ and $\vec{P}_{B}$, and is evaluated from the wave function overlaps between the initial and final states with the interaction potential $V_{ij}$.
This effective potential $V_{\rm eff}(\vec{P}_{\alpha},\vec{P}_{B})$ incorporates the interaction contributions from the four diagrams $C_{1}$, $C_{2}$, $T_{1}$, and $T_{2}$ shown in Fig.~\ref{fig3}.

For each diagram, $V_{eff}(\vec{P}_{\alpha},\vec{P}_{B})$ can be factorized into the product of three factors
\begin{eqnarray}
V_{{\rm eff}}(\vec{P}_{\alpha},\vec{P}_{B})
=I_{{\rm flavor}}I_{{\rm color}}I_{{\rm spin-space}}.
\end{eqnarray}
Here, $I$ with the subscripts flavor, color, and spin-space denotes the overlaps of the initial and final wave functions in the corresponding spaces.
The detailed derivations and numerical results of the above three overlaps are presented in Refs. \cite{Barnes:1991em,Wong:2001td,An:2025qfw,An:2025rjv}.
Finally, all possible partial decay widths are evaluated using Eq.~(\ref{width}) and summarized in Table~\ref{QQQq3}. 
Since certain suppressed contributions—such as three-body strong decays—are not included, the actual total widths are expected to be slightly larger than the values obtained in this work.

\section{Numerical analysis}\label{sec4}

\begin{table*}[t]
\centering
\caption{
Masses and internal energy contributions (in MeV) for different color–spin configurations of the triply heavy tetraquark systems $cc\bar{c}\bar{n}$,
$cc\bar{c}\bar{s}$, $bb\bar{b}\bar{n}$, and $bb\bar{b}\bar{s}$.
}\label{QQQq}
\begin{lrbox}{\tablebox}
\renewcommand\arraystretch{2.1}
\renewcommand\tabcolsep{1.2 pt}
\begin{tabular}{ccc|ccccc|cc|ccccc}
\toprule[1.50pt]
\toprule[0.50pt]
\multicolumn{3}{c|}{$cc\bar{c}\bar{n}$}&
\multicolumn{5}{c|}{Internal contribution}
&
\multicolumn{2}{c|}{$bb\bar{b}\bar{n}$}&
\multicolumn{5}{c}{Internal contribution}
\\
\Xcline{4-8}{0.3pt}
\Xcline{11-15}{0.3pt}
\multirow{1}*{$J^{P}$}
&\multirow{1}*{Configuration}
&\multirow{1}*{Mass}
&\multirow{1}*{$\langle T \rangle$}
&\multirow{1}*{$\langle V^{\rm C} \rangle$}
&\multirow{1}*{$\langle V^{\rm Coul} \rangle$}
&\multirow{1}*{$\langle V^{\rm Lin} \rangle$}
&\multirow{1}*{$\langle V^{\rm CS} \rangle$}
&\multirow{1}*{Configuration}
&\multirow{1}*{Mass}
&\multirow{1}*{$\langle T \rangle$}
&\multirow{1}*{$\langle V^{\rm C} \rangle$}
&\multirow{1}*{$\langle V^{\rm Coul} \rangle$}
&\multirow{1}*{$\langle V^{\rm Lin} \rangle$}
&\multirow{1}*{$\langle V^{\rm CS} \rangle$}
\\
\bottomrule[1.00pt]
\multirow{1}*{$2^{+}$}&\multirow{1}*{$
|(cc)^{\bar{3}_{c}}_{1}(\bar{c}\bar{n})^{3_{c}}_{1}\rangle_{2}$}&5386.3&734.8&-1657.3&-582.9&952.0&43.8
&\multirow{1}*{$
|(bb)^{\bar{3}_{c}}_{1}(\bar{b}\bar{n})^{3_{c}}_{1}\rangle_{2}$}&15147.8&770.0&-2146.5&-850.3&730.2&23.3
\\
\multirow{1}*{$1^{+}$}&\multirow{1}*{$
|(cc)^{\bar{3}_{c}}_{1}(\bar{c}\bar{n})^{3_{c}}_{1}\rangle_{1}$}
&5342.2&762.7&-1685.6&-594.1&934.9&0.2
&\multirow{1}*{$
|(bb)^{\bar{3}_{c}}_{1}(\bar{b}\bar{n})^{3_{c}}_{1}\rangle_{1}$}
&15122.8&786.8&-2163.4&-860.1&723.1&-1.6
\\
&\multirow{1}*{$
|(cc)^{\bar{3}_{c}}_{1}(\bar{c}\bar{n})^{3_{c}}_{0}\rangle_{1}$}
&5306.8&786.2&-1707.8&-601.7&920.3&-36.6
&\multirow{1}*{$
|(bb)^{\bar{3}_{c}}_{1}(\bar{b}\bar{n})^{3_{c}}_{0}\rangle_{1}$}
&15110.2&794.7&-2171.1&-862.1&717.4&-14.4
\\
&\multirow{1}*{$
|(cc)^{6_{c}}_{0}(\bar{c}\bar{n})^{\bar{6}_{c}}_{1}\rangle_{1}$}
&5307.1&764.1&-1725.1&-616.3&917.6&3.1
&\multirow{1}*{$
|(bb)^{6_{c}}_{0}(\bar{b}\bar{n})^{\bar{6}_{c}}_{1}\rangle_{1}$}
&15078.0&809.3&-2234.5&-917.5&709.5&2.2
\\
\multirow{1}*{$0^{+}$}&\multirow{1}*{$
|(cc)^{\bar{3}_{c}}_{1}(\bar{c}\bar{n})^{3_{c}}_{1}\rangle_{0}$}
&5319.6&777.3&-1699.9&-600.0&926.4&-22.9
&\multirow{1}*{$
|(bb)^{\bar{3}_{c}}_{1}(\bar{b}\bar{n})^{3_{c}}_{1}\rangle_{0}$}
&15110.0&795.4&-2171.9&-865.1&719.6&-14.5
\\
&\multirow{1}*{$
|(cc)^{6_{c}}_{0}(\bar{c}\bar{n})^{\bar{6}_{c}}_{0}\rangle_{0}$}
&5330.6&748.3&-1709.0&-610.4&927.8&26.3
&\multirow{1}*{$
|(bb)^{6_{c}}_{0}(\bar{b}\bar{n})^{\bar{6}_{c}}_{0}\rangle_{0}$}
&15089.5&801.4&-2226.5&-914.0&713.8&13.6
\\
\bottomrule[1pt]
\multicolumn{3}{c|}{$cc\bar{c}\bar{s}$}&
\multicolumn{5}{c|}{Internal contribution}
&
\multicolumn{2}{c|}{$bb\bar{b}\bar{s}$}&
\multicolumn{5}{c}{Internal contribution}
\\
\Xcline{4-8}{0.3pt}
\Xcline{11-15}{0.3pt}
\multirow{1}*{$J^{P}$}
&\multirow{1}*{Configuration}
&\multirow{1}*{Mass}
&\multirow{1}*{$\langle T \rangle$}
&\multirow{1}*{$\langle V^{\rm C} \rangle$}
&\multirow{1}*{$\langle V^{\rm Coul} \rangle$}
&\multirow{1}*{$\langle V^{\rm Lin} \rangle$}
&\multirow{1}*{$\langle V^{\rm CS} \rangle$}
&\multirow{1}*{Configuration}
&\multirow{1}*{Mass}
&\multirow{1}*{$\langle T \rangle$}
&\multirow{1}*{$\langle V^{\rm C} \rangle$}
&\multirow{1}*{$\langle V^{\rm Coul} \rangle$}
&\multirow{1}*{$\langle V^{\rm Lin} \rangle$}
&\multirow{1}*{$\langle V^{\rm CS} \rangle$}
\\
\bottomrule[1.00pt]
\multirow{1}*{$2^{+}$}&\multirow{1}*{$
|(cc)^{\bar{3}_{c}}_{1}(\bar{c}\bar{s})^{3_{c}}_{1}\rangle_{2}$}
&5489.4&723.9&-1754.9&-615.8&887.3&40.4
&\multirow{1}*{$
|(bb)^{\bar{3}_{c}}_{1}(\bar{b}\bar{s})^{3_{c}}_{1}\rangle_{2}$}
&15238.1&762.5&-2262.4&-898.1&662.1&22.0
\\
\multirow{1}*{$1^{+}$}&\multirow{1}*{$
|(cc)^{\bar{3}_{c}}_{1}(\bar{c}\bar{s})^{3_{c}}_{1}\rangle_{1}$}
&5449.9&749.1&-1780.5&-626.5&872.4&1.3
&\multirow{1}*{$
|(bb)^{\bar{3}_{c}}_{1}(\bar{b}\bar{s})^{3_{c}}_{1}\rangle_{1}$}
&15215.4&777.8&-2277.8&-907.3&656.0&-0.6
\\
&\multirow{1}*{$
|(cc)^{\bar{3}_{c}}_{1}(\bar{c}\bar{s})^{3_{c}}_{0}\rangle_{1}$}
&5417.9&770.0&-1800.3&-634.6&860.6&-31.8
&\multirow{1}*{$
|(bb)^{\bar{3}_{c}}_{1}(\bar{b}\bar{s})^{3_{c}}_{0}\rangle_{1}$}
&15202.6&785.8&-2285.6&-910.4&651.2&-13.6
\\
&\multirow{1}*{$
|(cc)^{6_{c}}_{0}(\bar{c}\bar{s})^{\bar{6}_{c}}_{1}\rangle_{1}$}
&5411.4&745.8&-1818.5&-645.3&853.2&4.1
&\multirow{1}*{$
|(bb)^{6_{c}}_{0}(\bar{b}\bar{s})^{\bar{6}_{c}}_{1}\rangle_{1}$}
&15170.7&798.6&-2346.4&-961.2&641.2&2.4
\\
\multirow{1}*{$0^{+}$}&\multirow{1}*{$
|(cc)^{\bar{3}_{c}}_{1}(\bar{c}\bar{s})^{3_{c}}_{1}\rangle_{0}$}
&5429.7&762.2&-1793.3&-631.9&865.1&-19.2
&\multirow{1}*{$
|(bb)^{\bar{3}_{c}}_{1}(\bar{b}\bar{s})^{3_{c}}_{1}\rangle_{0}$}
&15204.0&785.5&-2285.4&-912.0&652.9&-12.1
\\
&\multirow{1}*{$
|(cc)^{6_{c}}_{0}(\bar{c}\bar{s})^{\bar{6}_{c}}_{0}\rangle_{0}$}
&5432.1&731.5&-1803.9&-639.3&861.8&24.5
&\multirow{1}*{$
|(bb)^{6_{c}}_{0}(\bar{b}\bar{s})^{\bar{6}_{c}}_{0}\rangle_{0}$}
&15181.6&791.2&-2338.9&-957.3&644.8&13.3
\\
\bottomrule[0.50pt]
\bottomrule[1.50pt]
\end{tabular}
\end{lrbox}\scalebox{1}{\usebox{\tablebox}}
\end{table*}

The predicted masses corresponding to each configuration of the $cc\bar{c}\bar{q}$ and $bb\bar{b}\bar{q}$ systems are summarized in Table \ref{QQQq}.
A detailed analysis of the contributions from each component of the Hamiltonian (Eqs.~(\ref{Eq1}-\ref{Eq2})) to these configurations is performed, with the results also presented in Table~\ref{QQQq}.
The results indicate that the average kinetic energy $\langle T\rangle$, the Coulomb potential $\langle V^{\rm Coul}\rangle$, and the linear confining potential $\langle V^{\rm Lin}\rangle$ are of the same order of magnitude.
Furthermore, it is found that the contribution of the hyperfine interaction potential $\langle V^{\rm CS}\rangle$, which is proportional to ($1/m_{i}m_{j}$), 
has a magnitude of approximately 0–30 MeV. 
This suppression arises because the constituent masses of the charm and bottom quarks exceed 1 GeV, reducing the hyperfine contribution relative to the other terms.
Accordingly, the hyperfine contribution in the $bb\bar{b}\bar{q}$ system is smaller than that of the $cc\bar{c}\bar{q}$ system.


Although the hyperfine interaction $\langle V^{\rm CS}\rangle$ provides only a small contribution to the mass, its nonvanishing off-diagonal matrix elements play a crucial role in mixing the $|(QQ)^{\rm color_{1}}_{\rm spin_{1}}(\bar{Q}\bar{q})^{\rm color_{2}}_{\rm spin_{2}}\rangle_{\rm spin_{3}}$ configurations.
After considering the configuration mixing, we obtain the mass spectra of the corresponding physical states and summarize them in Table \ref{QQQq2}. 
According to Table \ref{QQQq2}, such color-spin configuration mixing leads to notable mass shifts of the physical states.
The mass gaps between the physical states become larger than those of their pre-mixing counterparts.
Compared with the $cc\bar{c}\bar{q}$ system, the $bb\bar{b}\bar{q}$ system exhibits a weaker degree of mixing, resulting in relatively smaller mass gaps.

According to Table \ref{QQQq2}, the relative mass positions of each state
are illustrated in Figs. \ref{fig-cccn}–\ref{fig-bbbn}.
The ground-state masses of the $cc\bar{c}\bar{q}$ and $bb\bar{b}\bar{q}$ systems lie in the range of 5.2-5.5 GeV and 15.0-15.3 GeV, respectively.
For clarity, we label all possible spin quantum numbers of the rearrangement decay channels with subscripts.
When the spin of an initial tetraquark state is equal to the number in the superscript of a meson-meson final state, 
the tetraquark state can decay into that meson-meson channel, which is allowed by the angular momentum conservation.
As shown in Figs. \ref{fig-cccn}–\ref{fig-bbbn}, there is 
no stable state in either the $cc\bar{c}\bar{q}$ or $bb\bar{b}\bar{q}$ system.
Instead, all states are unstable and undergo rearrangement decays into final states composed of a singly heavy meson and a heavy quarkonium.
The primary reason is that the pairwise attractive interactions provided by $\langle V^{\rm C} \rangle$ are significantly weaker than those within the corresponding singly-heavy meson and heavy quarkonia.

Since all predicted physical states lie above their corresponding meson–meson decay thresholds, as shown in Figs.~\ref{fig-cccn}-\ref{fig-bbbn},
we further calculate the rearrangement decay partial widths of each state, which are computed via Eq.~(\ref{width}).
According to Table \ref{QQQq3}, 
the total widths of $cc\bar{c}\bar{q}$ and $bb\bar{b}\bar{q}$ states are predicted to be in the range of 15-50 MeV.
For some narrow resonances, 
though the phase space is large, 
the Feynman amplitudes $\mathcal{M}(A\to BC)$ derived from the four distinct quark-interchange diagrams have opposite signs, leading to substantial mutual cancellation of their respective contributions.
This effect significantly suppresses the decay width, thereby accounting for the formation of these narrow states.
The radiative decay widths are also evaluated using Eq.~(\ref{rad4}). 
These transitions yield widths below 4keV and are negligible compared to the total widths.
For clarity, the rearrangement decay partial widths, radiative decay widths, and total decay widths of each state are also illustrated in Figs.~\ref{fig-cccn}-\ref{fig-bbbn}.
Notably, their rearrangement decay properties can play a crucial role in distinguishing these tetraquark states from partner states with similar masses.
We further extract the partial width ratios, which enable us to suggest promising experimental channels for searching for these tetraquark states in specific meson–meson final states.

In addition to the mass spectra, internal mass contributions, and decay properties, we also compute the RMS radii using Eq.~\ref{rms}, with the results summarized in Table~\ref{QQQq3}.
From Table~\ref{QQQq3}, our calculation yields the relations$\langle r^{2}_{13} \rangle^{\frac{1}{2}}=\langle r^{2}_{23} \rangle^{\frac{1}{2}}$, $\langle r^{2}_{14} \rangle^{\frac{1}{2}}=\langle r^{2}_{24} \rangle^{\frac{1}{2}}$, and $\langle r^{2}_{13-24} \rangle^{\frac{1}{2}}=\langle r^{2}_{14-23} \rangle^{\frac{1}{2}}$, which are fully consistent with the symmetry analysis discussed in Subsec.~\ref{secaa}.
Moreover, the interparticle RMS radii $\langle r^{2}_{i-j}\rangle^{1/2}$ are of the same order of magnitude: they span 0.4–0.6 fm for the $cc\bar{c}\bar{q}$ system and 0.2–0.5 fm for the $bb\bar{b}\bar{q}$ system.
The intersubcluster distances 
$\langle r^{2}_{i-j}\rangle^{1/2}$ are approximately 0.45 fm for the $cc\bar{c}\bar{q}$ system and 0.35 fm for the $bb\bar{b}\bar{q}$ system.
This implies that the average separation between the two subclusters is comparable to that between individual particles, indicating a strong spatial overlap between the subclusters.
For a purely molecular configuration, $\langle r^{2}_{13-24} \rangle^{\frac{1}{2}}$ and $\langle r^{2}_{14-23} \rangle^{\frac{1}{2}}$ would be considerably larger than the remaining RMS radii—in particular $\langle r^{2}_{12-34} \rangle^{\frac{1}{2}}$ and on the order of a few femtometers, with the spatial overlap components being negligible in comparison.
Our numerical results therefore align with the expectations for the compact tetraquark picture.

To ensure clarity in the subsequent discussion, we introduce the notation $\rm T_{content}(J^{P},Mass)$ to label a specific tetraquark state.

\subsection{The $cc\bar{c}\bar{n}$ and $cc\bar{c}\bar{s}$ systems}

\begin{table*}[t]
\centering
\caption{
Color-spin configuration mixing results for the triply heavy tetraquark systems $cc\bar{c}\bar{n}$,
$cc\bar{c}\bar{s}$, $bb\bar{b}\bar{n}$, and $bb\bar{b}\bar{s}$, including the mixing matrix elements $\langle H\rangle$, the mass spectra, and the corresponding eigenvectors.
}\label{QQQq2}
\begin{lrbox}{\tablebox}
\renewcommand\arraystretch{1.7}
\renewcommand\tabcolsep{4.5 pt}
\begin{tabular}{ccccccc}
\toprule[1.50pt]
\toprule[0.50pt]
\multicolumn{1}{c}{State}&\multirow{1}*{$J^{P}$}&\multirow{1}*{Configuration}&\multirow{1}*{$\langle H \rangle$}&\multirow{1}*{Mass}&\multirow{1}*{Eigenvector}\\
\bottomrule[1.00pt]
\multirow{6}*{$cc\bar{c}\bar{n}$}&
\multirow{1}*{$2^{+}$}&$|(cc)^{\bar{3}_{c}}_{1}(\bar{c}\bar{n})^{3_{c}}_{1}\rangle_{2}$&5386&5386&1
\\
&\multirow{3}*{$1^{+}$}&
\multirow{3}*{$\begin{pmatrix}
|(cc)^{\bar{3}_{c}}_{1}(\bar{c}\bar{n})^{3_{c}}_{1}\rangle_{1}\\
|(cc)^{\bar{3}_{c}}_{1}(\bar{c}\bar{n})^{3_{c}}_{0}\rangle_{1}\\
|(cc)^{6_{c}}_{0}(\bar{c}\bar{n})^{\bar{6}_{c}}_{1}\rangle_{1}
\end{pmatrix}$}
&
\multirow{3}*{$\begin{pmatrix}
5342&8&18\\
8&5307&-50\\
18&-50&5307\\
\end{pmatrix}$}
&
\multirow{3}*{$\begin{pmatrix}
5360\\5343\\5253
\end{pmatrix}$}
&
\multirow{3}*{$\begin{pmatrix}
0.43&-0.58&0.69\\
0.88&0.44&-0.18\\
0.20&-0.68&-0.70\\
\end{pmatrix}$}\\
\\
\\
&\multirow{2}*{$0^{+}$}&
\multirow{2}*{$\begin{pmatrix}
|(cc)^{\bar{3}_{c}}_{1}(\bar{c}\bar{n})^{3_{c}}_{1}\rangle_{0}\\
|(cc)^{6_{c}}_{0}(\bar{c}\bar{n})^{\bar{6}_{c}}_{0}\rangle_{0}
\end{pmatrix}$}
&
\multirow{2}*{$\begin{pmatrix}
5320&80\\
80&5331\\
\end{pmatrix}$}
&
\multirow{2}*{$\begin{pmatrix}
5405\\
5245\\
\end{pmatrix}$}&
\multirow{2}*{$\begin{pmatrix}
0.68&0.73\\
-0.73&0.68\\
\end{pmatrix}$}
\\
\\
\bottomrule[0.75pt]
\multirow{6}*{$cc\bar{c}\bar{s}$}&
\multirow{1}*{$2^{+}$}&$|(cc)^{\bar{3}_{c}}_{1}(\bar{c}\bar{s})^{3_{c}}_{1}\rangle_{2}$&5489&5489&1
\\
&\multirow{3}*{$1^{+}$}&
\multirow{3}*{$\begin{pmatrix}
|(cc)^{\bar{3}_{c}}_{1}(\bar{c}\bar{s})^{3_{c}}_{1}\rangle_{1}\\
|(cc)^{\bar{3}_{c}}_{1}(\bar{c}\bar{s})^{3_{c}}_{0}\rangle_{1}\\
|(cc)^{6_{c}}_{0}(\bar{c}\bar{s})^{\bar{3}_{c}}_{1}\rangle_{1}
\end{pmatrix}$}
&
\multirow{3}*{$\begin{pmatrix}
5450&5&12\\
5&5418&-45\\
12&-45&5411\\
\end{pmatrix}$}
&
\multirow{3}*{$\begin{pmatrix}
5461\\5450\\5368
\end{pmatrix}$}
&
\multirow{3}*{$\begin{pmatrix}
0.42&-0.63&0.66\\
0.90&0.40&-0.19\\
0.15&-0.67&-0.73\\
\end{pmatrix}$}\\
\\
\\
&\multirow{2}*{$0^{+}$}&
\multirow{2}*{$\begin{pmatrix}
|(cc)^{\bar{3}_{c}}_{1}(\bar{c}\bar{s})^{3_{c}}_{1}\rangle_{0}\\
|(cc)^{6_{c}}_{0}(\bar{c}\bar{s})^{\bar{3}_{c}}_{0}\rangle_{0}
\end{pmatrix}$}
&
\multirow{2}*{$\begin{pmatrix}
5430&71\\
71&5432\\
\end{pmatrix}$}
&
\multirow{2}*{$\begin{pmatrix}
5502\\
5360\\
\end{pmatrix}$}&
\multirow{2}*{$\begin{pmatrix}
0.70&0.71\\
-0.71&0.70\\
\end{pmatrix}$}
\\
\\
\bottomrule[0.75pt]
\multirow{6}*{$bb\bar{b}\bar{n}$}&
\multirow{1}*{$2^{+}$}&$|(bb)^{\bar{3}_{c}}_{1}(\bar{b}\bar{n})^{3_{c}}_{1}\rangle_{2}$&15148&15148&1
\\
&\multirow{3}*{$1^{+}$}&
\multirow{3}*{$\begin{pmatrix}
|(bb)^{\bar{3}_{c}}_{1}(\bar{b}\bar{n})^{3_{c}}_{1}\rangle_{1}\\
|(bb)^{\bar{3}_{c}}_{1}(\bar{b}\bar{n})^{3_{c}}_{0}\rangle_{1}\\
|(bb)^{6_{c}}_{0}(\bar{b}\bar{n})^{\bar{6}_{c}}_{1}\rangle_{1}
\end{pmatrix}$}
&
\multirow{3}*{$\begin{pmatrix}
15123&7&17\\
7&15110&-28\\
17&-28&15078\\
\end{pmatrix}$}
&
\multirow{3}*{$\begin{pmatrix}
15129\\15125\\15057
\end{pmatrix}$}
&
\multirow{3}*{$\begin{pmatrix}
0.80&-0.38&0.47\\
0.54&0.80&-0.28\\
-0.27&0.48&0.84\\
\end{pmatrix}$}\\
\\
\\
&\multirow{2}*{$0^{+}$}&
\multirow{2}*{$\begin{pmatrix}
|(bb)^{\bar{3}_{c}}_{1}(\bar{b}\bar{n})^{3_{c}}_{1}\rangle_{0}\\
|(bb)^{6_{c}}_{0}(\bar{b}\bar{n})^{\bar{6}_{c}}_{0}\rangle_{0}
\end{pmatrix}$}
&
\multirow{2}*{$\begin{pmatrix}
15110&47\\
47&15090\\
\end{pmatrix}$}
&
\multirow{2}*{$\begin{pmatrix}
15148\\
15052\\
\end{pmatrix}$}&
\multirow{2}*{$\begin{pmatrix}
-0.78&-0.63\\
0.63&-0.78\\
\end{pmatrix}$}
\\
\\
\bottomrule[0.75pt]
\multirow{6}*{$bb\bar{b}\bar{s}$}&
\multirow{1}*{$2^{+}$}&$|(bb)^{\bar{3}_{c}}_{1}(\bar{b}\bar{s})^{3_{c}}_{1}\rangle_{2}$&15238&15238&1
\\
&\multirow{3}*{$1^{+}$}&
\multirow{3}*{$\begin{pmatrix}
|(bb)^{\bar{3}_{c}}_{1}(\bar{b}\bar{s})^{3_{c}}_{1}\rangle_{1}\\
|(bb)^{\bar{3}_{c}}_{1}(\bar{b}\bar{s})^{3_{c}}_{0}\rangle_{1}\\
|(bb)^{6_{c}}_{0}(\bar{b}\bar{s})^{\bar{6}_{c}}_{1}\rangle_{1}
\end{pmatrix}$}
&
\multirow{3}*{$\begin{pmatrix}
15215&6&13\\
6&15203&-25\\
13&-25&15171\\
\end{pmatrix}$}
&
\multirow{3}*{$\begin{pmatrix}
15219\\15216\\15154
\end{pmatrix}$}
&
\multirow{3}*{$\begin{pmatrix}
0.89&-0.26&0.37\\
0.40&0.85&-0.35\\
0.22&-0.46&-0.86\\
\end{pmatrix}$}\\
\\
\\
&\multirow{2}*{$0^{+}$}&
\multirow{2}*{$\begin{pmatrix}
|(bb)^{\bar{3}_{c}}_{1}(\bar{b}\bar{s})^{3_{c}}_{1}\rangle_{0}\\
|(bb)^{6_{c}}_{0}(\bar{b}\bar{s})^{\bar{6}_{c}}_{0}\rangle_{0}
\end{pmatrix}$}
&
\multirow{2}*{$\begin{pmatrix}
15204&42\\
42&15182\\
\end{pmatrix}$}
&
\multirow{2}*{$\begin{pmatrix}
15236\\
15149\\
\end{pmatrix}$}&
\multirow{2}*{$\begin{pmatrix}
-0.79&-0.61\\
0.61&-0.79\\
\end{pmatrix}$}
\\
\\
\bottomrule[0.50pt]
\bottomrule[1.50pt]
\end{tabular}
\end{lrbox}\scalebox{1.0}{\usebox{\tablebox}}
\end{table*}

Starting from the results presented in Table~\ref{QQQq3} and Fig.~\ref{fig-cccn}, we first analyze the $cc\bar{c}\bar{n}$ and $cc\bar{c}\bar{s}$ systems.
There are two $J^{P}=0^{+}$ states, three $J^{P}=1^{+}$ states, and one $J^{P}=2^{+}$ state in the $cc\bar{c}\bar{n}$ and $cc\bar{c}\bar{s}$ systems.

Owing to the larger constituent quark mass of the strange quark, all quantum states in the $cc\bar{c}\bar{s}$ system are systematically heavier than their counterparts in the $cc\bar{c}\bar{n}$ system. 
For instance, the mass of the $J^{P}=2^{+}$ ground state increases from 5386 MeV (for the $cc\bar{c}\bar{n}$ state) to 5489 MeV (for the $cc\bar{c}\bar{s}$ state). 
On the other hand, since the $V^{CS}$ interaction (Eq. (\ref{Eq2}))— 
which dominates the mass gap—scales with $1/m_{i}m_{j}$, 
the magnitude of the mass gap between the corresponding states is relatively small. 
As an illustration, the mass gaps between the $J^{P}=1^{+}$ ground states decrease from 17 MeV and 90 MeV in the $cc\bar{c}\bar{n}$ system to 11 MeV and 82 MeV in the $cc\bar{c}\bar{s}$ system.

It should be noted that although the $cc\bar{c}\bar{u}/cc\bar{c}\bar{d}$ and $bb\bar{b}\bar{u}/bb\bar{b}\bar{d}$ states exhibit identical mass spectra 
(when neglecting the small difference in the constituent quark masses of the $\bar{u}$ and $\bar{d}$ quarks), 
the distinct electric charges of the $\bar{u}$ and $\bar{d}$ valence quarks lead to different transition magnetic moments among these states, 
which in turn result in different radiative decay widths. 
Accordingly, we present the distinct numerical results for the radiative decay widths of the $cc\bar{c}\bar{u}/cc\bar{c}\bar{d}$ and $bb\bar{b}\bar{u}/bb\bar{b}\bar{d}$ states in Table~\ref{QQQq3} and Figs.~\ref{fig-cccn}-\ref{fig-bbbn}. 
Furthermore, compared to the radiative decay widths, two-body strong decays are the dominant decay channels for these resonances. 
Consequently, the $cc\bar{c}\bar{u}/cc\bar{c}\bar{d}$ and $bb\bar{b}\bar{u}/bb\bar{b}\bar{d}$ states have nearly identical total decay widths.

First, for the $J^{P}=2^{+}$ $cc\bar{c}\bar{s}$ state, $T_{c^{2}\bar{c}\bar{s}}(5489,2^{+})$, predominantly decays into the $J/\psi D_{s}^{*}$ final state through an $S$-wave  rearrangement process, with a partial width of 24 MeV.
It can also decay into the $\eta_{c} D_{s}$ final states via a $D$-wave, though this channel is significantly suppressed.
In the radiative decay channel, most of the transitions have relatively small partial widths, typically less than 2 keV.
Similarly, owing to angular momentum conservation, the state with $J^{P}=2^{+}$ cannot undergo a radiative transition to the $J^{P}=0^{+}$ state.
Notably, the radiative transition of $T_{c^{2}\bar{c}\bar{s}}(5489,2^{+})$ to $T_{c^{2}\bar{c}\bar{s}}(5368,1^{+})$ amounts to 1.4 keV, whereas the transitions to $T_{c^{2}\bar{c}\bar{s}}(5461,1^{+})$ and $T_{c^{2}\bar{c}\bar{s}}(5450,1^{+})$ are negligible.
This narrow total width, dominated by a single strong decay channel, suggests that the $J^{P}=2^{+}$
state could appear as a clean resonance in the $J/\psi D_{s}^{*}$ invariant mass distribution.

For three $J^{P}=1^{+}$ $cc\bar{c}\bar{s}$ states, each has three distinct rearrangement decay channels: $J/\psi D^{*}_{s}$, $J/\psi D_{s}$, and $\eta_{c} D^{*}_{s}$.
Among these, $T_{c^{2}\bar{c}\bar{s}}(5368,1^{+})$ has the lowest mass and the largest total decay width of approximately 34 MeV.
Its corresponding relative partial decay width ratio is as follows:
\begin{eqnarray}
\Gamma_{J/\psi D_{s}^{*}}:\Gamma_{J/\psi D_{s}}:\Gamma_{\eta_{c} D_{s}^{*}}=3:1:2.2.
\end{eqnarray}
From the above ratio, we notice that $J/\psi D_{s}^{*}$ channel is the dominant decay channel for $T_{c^{2}\bar{c}\bar{s}}(5368,1^{+})$.
The other two states are partner states, with their mass gap being only 10 MeV. 
Given theoretical uncertainties and experimental precision,
it is challenging to distinguish them experimentally based on mass alone.
However, their total decay widths and relative branching ratios exhibit significant differences. 
Specifically, the total width of $T_{c^{2}\bar{c}\bar{s}}(5461,1^{+})$ is 31 MeV, while that of $T_{c^{2}\bar{c}\bar{s}}(5450,1^{+})$ is 23 MeV. 
Accordingly, the latter manifests as a narrower resonance, whose spectral peak can be clearly resolved from the background. 
Furthermore, their relative partial decay width ratios:
\begin{eqnarray}
\Gamma_{J/\psi D_{s}^{*}}:\Gamma_{J/\psi D_{s}}:\Gamma_{\eta_{c} D_{s}^{*}}&=&2.4:1:2.2,\nonumber\\
\Gamma_{J/\psi D_{s}^{*}}:\Gamma_{J/\psi D_{s}}:\Gamma_{\eta_{c} D_{s}^{*}}&=&1:2:4.2,
\end{eqnarray}
correspond to $T_{c^{2}\bar{c}\bar{s}}(5461,1^{+})$ and $T_{c^{2}\bar{c}\bar{s}}(5450,1^{+})$, respectively.
From the above ratios, we notice that the $J/\psi D^{*}_{s}$ channel is the dominant decay channel for $T_{c^{2}\bar{c}\bar{s}}(5461,1^{+})$.
In contrast, the $J/\psi D^{*}_{s}$ channel is suppressed for $T_{c^{2}\bar{c}\bar{s}}(5450,1^{+})$, 
which decays predominantly to the $\eta_{c} D_{s}^{*}$ final state.
On the other hand, these two partner states exhibit drastically different radiative transition strengths toward the ground state $T_{c^{2}\bar{c}\bar{s}}(5368,1^{+})$, which serves as an efficient discriminative observable. 
The radiative decay width of $T_{c^{2}\bar{c}\bar{s}}(5461,1^{+})\to T_{c^{2}\bar{c}\bar{s}}(5368,1^{+})$ reaches 0.78 keV, which is sufficiently large to be potentially accessible in high-precision measurements. 
In sharp contrast, the corresponding radiative transition for $T_{c^{2}\bar{c}\bar{s}}(5450,1^{+})\to T_{c^{2}\bar{c}\bar{s}}(5368,1^{+})$ is only 0.09 keV. 
Such a significant discrepancy in radiative decay strengths offers a unique experimental criterion. If a measurable radiative transition signal toward the $T_{c^{2}\bar{c}\bar{s}}(5368,1^{+})$ is observed experimentally, the corresponding resonance can be unambiguously attributed to the $T_{c^{2}\bar{c}\bar{s}}(5461,1^{+})$ state rather than its near-degenerate partner.

For two $J^{P}=0^{+}$ $cc\bar{c}\bar{s}$ states, 
$T_{c^{2}\bar{c}\bar{s}}(5502,0^{+})$ is the highest-mass tetraquark state in the $cc\bar{c}\bar{s}$ system. 
Its sufficiently large decay phase space allows three feasible radiative decay pathways.
Numerical results indicate that the radiative transition toward the $T_{c^2\bar{c}\bar{s}}(5368,1^+)$ dominates its radiative decay behaviors, with a tiny decay width of approximately 1 keV. 
Compared with strong decay processes, the contribution of radiative transitions is negligible and barely alters the resonance profile and overall decay properties. 
Accordingly, the observable characteristics of $T_{c^{2}\bar{c}\bar{s}}(5502,0^{+})$ are still predominantly governed by the rearrangement decay mechanism.
Within the quark-exchange rearrangement framework, 
the$T_{c^2\bar{c}\bar{s}}(5502,0^+)$ state decays into the$J/\psi D_s^*$ and $\eta_c D_s$ final states. 
The corresponding partial decay width ratio reads
\begin{eqnarray}
\frac{\Gamma[T_{c^{2}\bar{c}\bar{s}}(5502,0^{+})\to J/\psi D_{s}^{*}]}{\Gamma[T_{c^{2}\bar{c}\bar{s}}(5502,0^{+})\to \eta_{c} D_{s}]}=1.5,
\end{eqnarray}
indicating that these two decay modes contribute comparably and jointly dominate the strong decay behaviors of this resonance. 
This state has a total strong decay width of 19 MeV, 
corresponding to a typical narrow resonance with a well-defined peak structure and negligible background contamination, which favors future experimental observations.
The other $0^+$ state $T_{c^2\bar{c}\bar{s}}(5360,0^+)$ also exhibits prominent narrow-resonance behavior, with a total decay width below 20 MeV. 
Meanwhile, its relative partial decay width ratio is:
\begin{eqnarray}
\frac{\Gamma[T_{c^{2}\bar{c}\bar{s}}(5360,0^{+})\to J/\psi D_{s}^{*}]}{\Gamma[T_{c^{2}\bar{c}\bar{s}}(5360,0^{+})\to \eta_{c} D_{s}]}=1.5.
\end{eqnarray}
Thus, both $J/\psi D_{s}^{*}$ and $\eta_{c} D_{s}$ channels are its main decay channels.
Therefore, we suggest that experiments scan the invariant mass spectra of $J/\psi D^{*}_{s}$ or $\eta_{c} D_{s}$ in the mass range of 5.3-5.4 GeV, and the branching ratio conforms to the 1:1.5 ratio, with particular attention paid to the narrow peak structure near 5360 MeV.
If the peak position and width observed in the experiment are consistent with
theoretical predictions, this peak can be confirmed as the signal of $T_{c^{2}\bar{c}\bar{s}}(5360,0^{+})$.
Moreover, based on the above analysis of the $cc\bar{c}\bar{s}$ states, and guided by the results presented in Table \ref{QQQq3} and Fig. \ref{fig-cccn}, one can perform analogous discussions on the decay behaviors of $cc\bar{c}\bar{n}$ system, and further explore their characteristics in-depth.

\begin{table*}[t]
\centering
\caption{
RMS radii (in fm), radiative decay widths (in keV), rearrangement decay partial widths (in MeV), and total decay widths (in MeV) for the triply heavy tetraquark systems $cc\bar{c}\bar{n}$,
$cc\bar{c}\bar{s}$, $bb\bar{b}\bar{n}$, and $bb\bar{b}\bar{s}$.
For the $cc\bar{c}\bar{n}$ and $bb\bar{b}\bar{n}$ tetraquark states, e.g. 0.5/21 represents the radiative decay widths and total decay widths of the $cc\bar{c}\bar{u}$ ($bb\bar{b}\bar{u}$) and $cc\bar{c}\bar{d}$ ($bb\bar{b}\bar{d}$) states, respectively.
}\label{QQQq3}
\begin{lrbox}{\tablebox}
\renewcommand\arraystretch{2.3}
\renewcommand\tabcolsep{0.5 pt}
\begin{tabular}{cc|cccccc|cccc|cccc|c}
\toprule[1.50pt]
\toprule[0.50pt]
\multicolumn{2}{l|}{$cc\bar{c}\bar{n}$}&
\multicolumn{6}{c|}{RMS Radius}&\multicolumn{4}{c|}{Radiative decay properties}&\multicolumn{5}{r}{Rearrangement decay properties}\\
\Xcline{3-17}{0.3pt}
\multirow{2}*{$J^{P}$}&\multirow{2}*{State}
&\multirow{2}*{$R_{12}$}&\multirow{2}*{$R_{34}$}
&$R_{13}$&$R_{14}$
&\multirow{2}*{$R_{12-34}$}
&$R_{13-24}$
&\multirow{2}*{$T(5360,1^{+})\gamma$}
&\multirow{2}*{$T(5343,1^{+})\gamma$}
&\multirow{2}*{$T(5253,1^{+})\gamma$}
&\multirow{2}*{$T(5245,0^{+})\gamma$}
&\multirow{2}*{$J/\psi D^{*}$}
&\multirow{2}*{$J/\psi D$}
&\multirow{2}*{$\eta_{c} D^{*}$}
&\multirow{2}*{$\eta_{c} D$}
&\multirow{2}*{$\Gamma_{sum}$}\\
&&&&$R_{23}$&$R_{24}$&&$R_{14-23}$&&&&&&&&&\\
\bottomrule[0.50pt]
$2^{+}$&$T_{c^{2}\bar{c}\bar{n}}(5386,2^{+})$&0.42&0.59&0.47&0.50&0.48&0.47&\scriptsize{$6.6\times10^{-4}$}$|$\scriptsize{$2.1\times10^{-2}$}&0.78$|$$1.4\times10^{-2}$&0.71$|$2.9&&47.0&&&&47.0
\\
\multirow{1}*{$1^{+}$}&$T_{c^{2}\bar{c}\bar{n}}(5360,1^{+})$&
0.46&0.60&0.44&0.47&0.41&0.49&&\scriptsize{$4.3\times10^{-2}$}$|$\scriptsize{$1.2\times10^{-3}$}&3.2$|$1.1&0.53$|$0.36&18.9&9.2&13.7&&41.8
\\
&$T_{c^{2}\bar{c}\bar{n}}(5343,1^{+})$&
0.39&0.56&0.50&0.52&0.53&0.45&&&2.4$|$0.27&2.9$|$0.11&
5.2&13.9&21.5&&40.6
\\
&$T_{c^{2}\bar{c}\bar{n}}(5253,1^{+})$&
0.46&0.60&0.44&0.47&0.41&0.49&&&&\scriptsize{$1.7\times10^{-3}$}$|$\scriptsize{$8.0\times10^{-4}$}&22.3&8.0&6.1&&36.4
\\
\multirow{1}*{$0^{+}$}&$T_{c^{2}\bar{c}\bar{n}}(5405,0^{+})$&
0.45&0.61&0.47&0.50&0.46&0.49&1.2$|$0.31&1.2$|$0.37&$7.7\times10^{-3}$$|$1.9&&24.5&&&14.7&39.2
\\
&$T_{c^{2}\bar{c}\bar{n}}(5245,0^{+})$&
0.43&0.57&0.44&0.46&0.42&0.47&&&&&22.9&&&15.9&38.8
\\
\bottomrule[1.00pt]
\multicolumn{2}{l}{$cc\bar{c}\bar{s}$}&\multicolumn{1}{c}{}&\multicolumn{1}{c}{}&\multicolumn{1}{c}{}&&&
&\multirow{1}*{$T(5461,1^{+})\gamma$}
&\multirow{1}*{$T(5450,1^{+})\gamma$}
&\multirow{1}*{$T(5368,1^{+})\gamma$}
&\multirow{1}*{$T(5360,0^{+})\gamma$}
&\multirow{1}*{$J/\psi D_{s}^{*}$}
&\multirow{1}*{$J/\psi D_{s}$}
&\multirow{1}*{$\eta_{c} D_{s}^{*}$}
&\multirow{1}*{$\eta_{c} D_{s}$}
&\multicolumn{1}{c}{\multirow{1}*{$\Gamma_{sum}$}}\\
\bottomrule[0.50pt]
$2^{+}$&$T_{c^{2}\bar{c}\bar{s}}(5489,2^{+})$&
0.41&0.53&0.46&0.47&0.47&0.45&$2.4\times10^{-2}$&$1.6\times10^{-5}$&1.4&&23.6&&&&23.6
\\
$1^{+}$&$T_{c^{2}\bar{c}\bar{s}}(5461,1^{+})$&
0.43&0.54&0.45&0.46&0.44&0.47&&$2.8\times10^{-4}$&0.78&0.22&13.6&5.6&12.2&&31.4
\\
&$T_{c^{2}\bar{c}\bar{s}}(5450,1^{+})$&
0.39&0.50&0.48&0.49&0.50&0.43&&&$9.2\times10^{-2}$&0.15&3.2&6.7&13.4&&23.3
\\
&$T_{c^{2}\bar{c}\bar{s}}(5368,1^{+})$&
0.45&0.54&0.43&0.44&0.39&0.48&&&&$5.1\times10^{-4}$&16.2&5.4&11.9&&33.5
\\
$0^{+}$&$T_{c^{2}\bar{c}\bar{s}}(5502,0^{+})$&
0.44&0.54&0.45&0.47&0.44&0.48&0.11&0.26&1.0&&11.3&&&7.3&18.6
\\
&$T_{c^{2}\bar{c}\bar{s}}(5360,0^{+})$&
0.43&0.52&0.43&0.44&0.40&0.4&&&&&11.1&&&7.4&18.5
\\ 
\bottomrule[1.00pt]
\multicolumn{2}{l}{$bb\bar{b}\bar{n}$}&\multicolumn{1}{c}{}&\multicolumn{1}{c}{}&\multicolumn{1}{c}{}&&&
&\multirow{1}*{$T(15129,1^{+})\gamma$}
&\multirow{1}*{$T(15125,1^{+})\gamma$}
&\multirow{1}*{$T(15057,1^{+})\gamma$}
&\multirow{1}*{$T(15052,0^{+})\gamma$}
&\multirow{1}*{$\Upsilon B^{*}$}
&\multirow{1}*{$\Upsilon B$}
&\multirow{1}*{$\eta_{b} B^{*}$}
&\multirow{1}*{$\eta_{b} B$}
&\multicolumn{1}{c}{\multirow{1}*{$\Gamma_{sum}$}}\\
\bottomrule[0.50pt]
$2^{+}$&$T_{b^{2}\bar{b}\bar{n}}(15148,2^{+})$&0.25&0.54&0.30&0.37&0.30&0.30
&\scriptsize{$3.1\times10^{-4}$}$|$\scriptsize{$1.1\times10^{-3}$}&\scriptsize{$7.4\times10^{-2}$}$|$\scriptsize{$2.0\times10^{-2}$}&0.38$|$$3.2\times10^{-2}$&&19.8&&&&19.8
\\
$1^{+}$&$T_{b^{2}\bar{b}\bar{n}}(15129,1^{+})$&0.29&0.53&0.29&0.35&0.27&0.32&
&\scriptsize{$4.2\times10^{-5}$}$|$\scriptsize{$5.2\times10^{-6}$}&\scriptsize{$4.0\times10^{-2}$}$|$\scriptsize{$2.3\times10^{-3}$}&\scriptsize{$2.9\times10^{-2}$}$|$\scriptsize{$5.1\times10^{-2}$}&3.8&7.3&10.0&&21.1
\\
&$T_{b^{2}\bar{b}\bar{n}}(15125,1^{+})$&0.22&0.51&0.34&0.40&0.36&0.30&
&&2.2$|$1.1&\scriptsize{$4.3\times10^{-2}$}$|$\scriptsize{$3.2\times10^{-2}$}&7.0&5.7&7.9&&20.6
\\
&$T_{b^{2}\bar{b}\bar{n}}(15057,1^{+})$&0.29&0.53&0.29&0.35&0.27&0.32&
&&&\scriptsize{$5.9\times10^{-4}$}$|$\scriptsize{$1.3\times10^{-4}$}&9.9&5.0&6.4&&21.3
\\
$0^{+}$&$T_{b^{2}\bar{b}\bar{n}}(15148,0^{+})$&0.26&0.53&0.31&0.37&0.31&0.31&\scriptsize{$8.2\times10^{-2}$}$|$\scriptsize{$2.7\times10^{-2}$}&\scriptsize{$1.1\times10^{-2}$}$|$\scriptsize{$1.1\times10^{-4}$}&0.33$|$$1.2\times10^{-2}$&&9.6&&&12.5&22.1
\\
&$T_{b^{2}\bar{b}\bar{n}}(15052,0^{+})$&0.28&0.52&0.29&0.35&0.28&0.32&
&&&&12.2&&&10.0&22.2
\\
\bottomrule[1.00pt]
\multicolumn{2}{l}{$bb\bar{b}\bar{s}$}&\multicolumn{1}{c}{}&\multicolumn{1}{c}{}&\multicolumn{1}{c}{}&&&
&\multirow{1}*{$T(15219,1^{+})\gamma$}
&\multirow{1}*{$T(15216,1^{+})\gamma$}
&\multirow{1}*{$T(15154,1^{+})\gamma$}
&\multirow{1}*{$T(15149,0^{+})\gamma$}
&\multirow{1}*{$\Upsilon B_{s}^{*}$}
&\multirow{1}*{$\Upsilon B_{s}$}
&\multirow{1}*{$\eta_{b} B_{s}^{*}$}
&\multirow{1}*{$\eta_{b} B_{s}$}
&\multicolumn{1}{c}{\multirow{1}*{$\Gamma_{sum}$}}\\
\bottomrule[0.50pt]
$2^{+}$&$T_{b^{2}\bar{b}\bar{s}}(15238,2^{+})$&
0.24&0.46&0.30&0.34&0.30&0.30&$1.6\times10^{-3}$&$7.8\times10^{-3}$&$1.1\times10^{-2}$&&14.7&&&&14.7
\\
$1^{+}$&$T_{b^{2}\bar{b}\bar{s}}(15219,1^{+})$&
0.25&0.45&0.30&0.34&0.30&0.30&&$5.3\times10^{-6}$&$7.0\times10^{-5}$&$3.5\times10^{-2}$&2.4&6.3&8.4&&17.1
\\
&$T_{b^{2}\bar{b}\bar{s}}(15216,1^{+})$&
0.22&0.44&0.33&0.37&0.35&0.29&&&$5.0\times10^{-2}$&$9.4\times10^{-3}$&7.9&4.4&5.6&&17.9
\\
&$T_{b^{2}\bar{b}\bar{s}}(15154,1^{+})$&
0.29&0.46&0.28&0.32&0.26&0.32&&&&$4.0\times10^{-5}$&11.2&5.1&3.3&&19.6
\\
$0^{+}$&$T_{b^{2}\bar{b}\bar{s}}(15236,0^{+})$&
0.26&0.45&0.30&0.34&0.30&0.30&$1.0\times10^{-2}$&$4.0\times10^{-4}$&$4.4\times10^{-3}$&&8.9&&&6.4&15.3
\\
&$T_{b^{2}\bar{b}\bar{s}}(15149,0^{+})$&
0.28&0.45&0.28&0.32&0.26&0.31&&&&&13.7&&&6.2&19.9
\\
\bottomrule[0.50pt]
\bottomrule[1.50pt]
\end{tabular}
\end{lrbox}\scalebox{0.8}{\usebox{\tablebox}}
\end{table*}

\subsection{The $bb\bar{b}\bar{n}$ and $bb\bar{b}\bar{s}$ systems}

\begin{figure*}[htbp]
\begin{tabular}{c}
\includegraphics[width=440pt]{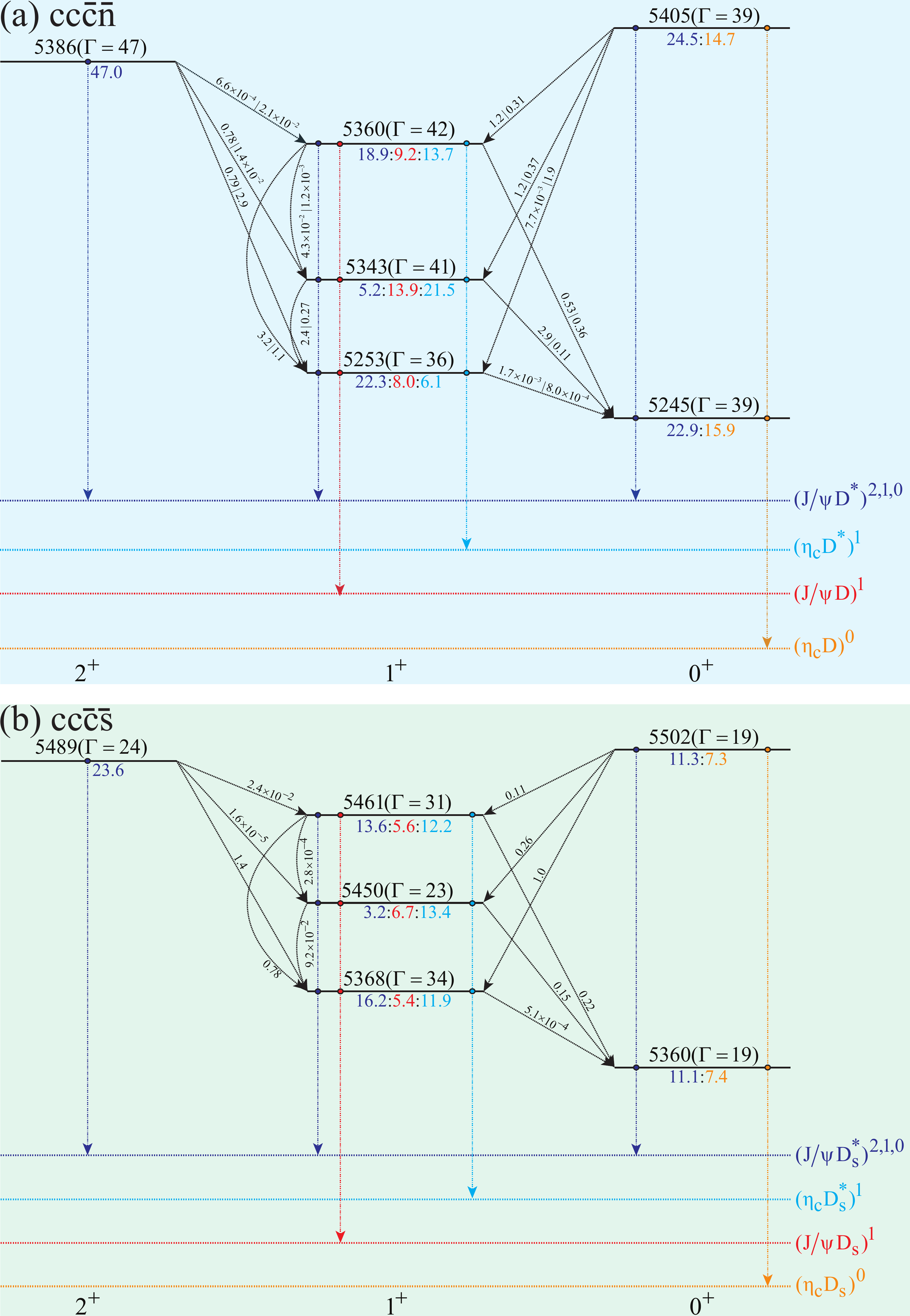}\\
\end{tabular}
\caption{
Relative positions for the $cc\bar{c}\bar{n}$ (a) and $cc\bar{c}\bar{s}$ (b) tetraquark states labeled with horizontal solid lines, e.g. $5489(\Gamma=24)$ represents the mass and total decay width of the corresponding state (units: MeV).
The numbers below the horizontal lines, e.g. $13.6:5.6:12.2$, represent the rearrangement decay partial widths of the corresponding state (units: MeV).
The dotted lines denote various $S$-wave meson-meson thresholds, and the superscripts of the labels, e.g. $(J/\psi D^{*})^{2,1,0}$, represent the possible total angular momenta of the channels.
The solid dots of different colors where the vertical dashed lines with arrows intersect the horizontal solid lines represent the allowed rearranged $S$-wave decay processes.
The black dashed lines represent radiative transitions between different states, with the adjacent numbers denoting the corresponding radiative decay widths ( units: keV). 
For the $cc\bar{c}\bar{n}$ (a) tetraquark state, e.g. $0.79|2.9$ represents the radiative decay widths of the $cc\bar{c}\bar{u}$ and $cc\bar{c}\bar{d}$ states, respectively, and the notation for their corresponding total widths follows the same convention.
}\label{fig-cccn}
\end{figure*}

\begin{figure*}[htbp]
\begin{tabular}{c}
\includegraphics[width=440pt]{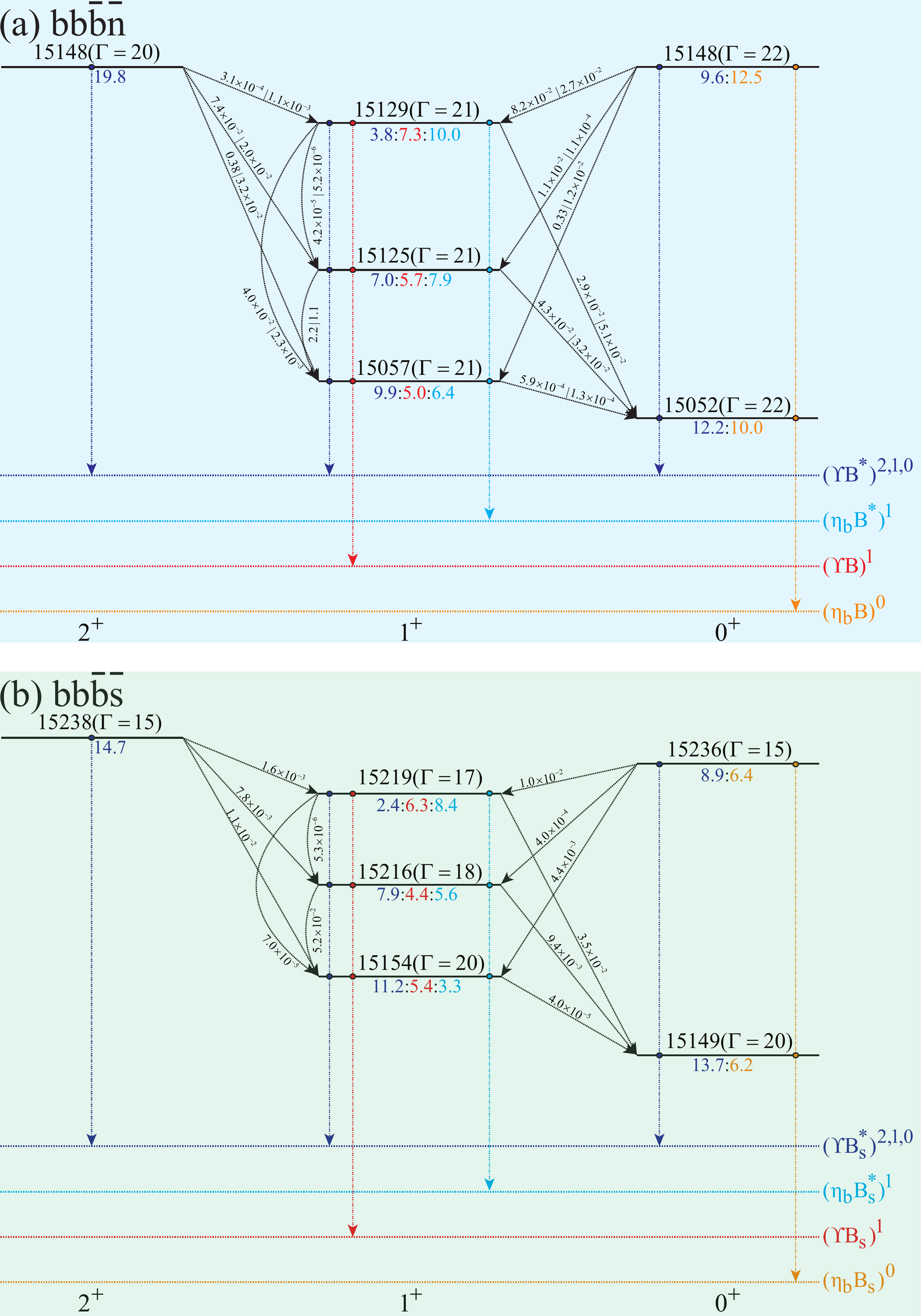}\\
\end{tabular}
\caption{
Relative positions for the $bb\bar{b}\bar{n}$ (a) and $bb\bar{b}\bar{s}$ (b) tetraquark states labeled with horizontal solid lines, e.g. $15238(\Gamma=15)$ represents the mass and total decay width of the corresponding state (units: MeV).
The numbers below the horizontal lines, e.g. $2.4:6.3:8.4$, represent the rearrangement decay partial widths of the corresponding state (units: MeV).
The dotted lines denote various $S$-wave meson-meson thresholds, and the superscripts of the labels, e.g. $(\Upsilon B^{*})^{2,1,0}$, represent the possible total angular momenta of the channels.
The solid dots of different colors where the vertical dashed lines with arrows intersect the horizontal solid lines represent the allowed rearranged $S$-wave decay processes.
The black dashed lines represent radiative transitions between different states, with the adjacent numbers denoting the corresponding radiative decay widths ( units: keV). 
For the $bb\bar{b}\bar{n}$ (a) tetraquark state, e.g. $3.1\times10^{-4}|1.1\times10^{-3}$ represents the radiative decay widths of the $bb\bar{b}\bar{u}$ and $bb\bar{b}\bar{d}$ states, respectively, and the notation for their corresponding total widths follows the same convention.
}\label{fig-bbbn}
\end{figure*}

Next, we discuss the $bb\bar{b}\bar{n}$ and $bb\bar{b}\bar{s}$ systems.
Since they obey identical symmetry constraints to the $cc\bar{c}\bar{n}$ and $cc\bar{c}\bar{s}$ systems, the number of allowed states is also identical.
As illustrated in Figs. \ref{fig-cccn} and \ref{fig-bbbn},
the $cc\bar{c}\bar{q}$ and $bb\bar{b}\bar{q}$ systems exhibit analogous mass spectra. 
This arises from the heavy quark symmetry, which becomes exact in the limit $m_{Q}\to\infty$, leading to identical symmetry constraints for both systems.
Further, according to heavy quark spin symmetry (HQSS), compared to the $cc\bar{c}\bar{q}$ system, the masses of hadrons with the same light degrees of freedom and spin should approach each other as the heavy quark mass increases in the $bb\bar{b}\bar{q}$ system, resulting in smaller mass gaps.
In contrast to the $cc\bar{c}\bar{q}$ system, the radiative decay widths of the $bb\bar{b}\bar{q}$ states are relatively smaller.
This is because the radiative transition amplitude $\mathcal{M}_{M_{J_f}, M_{J_i}}$ is proportional to $1/m_{q}$ ($m_{q}$=$m_{n}$,$m_{s}$,$m_{c}$, or $m_{b}$ denotes the constituent quark mass), and the much larger mass of the b-quark (relative to c-quark) suppresses radiative processes. 
For the radiative decay channels, all associated transitions have rather small partial widths, typically below 2.5 keV. 
In addition, the radiative decay widths of the $bb\bar{b}\bar{u}$ states are systematically larger than those of the $bb\bar{b}\bar{d}$ states, originating from subtle differences in light-quark electric charge.

Owing to angular momentum conservation, the $J^{P}=2^{+}$ $bb\bar{b}\bar{n}$ state $T_{b^{2}\bar{b}\bar{n}}(15148,2^{+})$ predominantly decays to the $\Upsilon B^{*}$ final state through an $S$-wave rearrangement process, with a partial width of 20 MeV for this channel.
Additionally, it can undergo $D$-wave transitions to the $\eta_{b} B$ final state, though this decay mode is strongly suppressed.
The radiative transition of $T_{b^{2}\bar{b}\bar{u}}(15148,2^{+})\to T_{b^{2}\bar{b}\bar{u}}(15057,1^{+})\gamma$ has a width of 0.38 keV,
and that to $T_{b^{2}\bar{b}\bar{u}}(15125,1^{+})$ is $7.4\times10^{-2}$ keV.
For the $d$-flavored state, the analogous transitions are only about 
$3.2\times10^{-2}$ keV and $2.0\times10^{-2}$ keV.
As a result, the total decay width of $T_{b^{2}\bar{b}\bar{u}}(15148,2^{+})$ is slightly larger than that of $T_{b^{2}\bar{b}\bar{d}}(15148,2^{+})$.

Among the three $J^{P}=1^{+}$ $bb\bar{b}\bar{n}$ states, $T_{b^{2}\bar{b}\bar{n}}(15129,\\1^{+})$ and $T_{b^{2}\bar{b}\bar{n}}(15125,1^{+})$ are partner states
with identical quantum numbers and similar masses.
Their mass gap is only 4 MeV, which is quite small compared with the significant difference in their total decay widths both are around 21 MeV, respectively.
Meanwhile, the $\eta_{b} B^{*}$ decay channel is the dominant decay mode for both states.
A distinct difference emerges between these two states: $T_{b^{2}\bar{b}\bar{n}}(15129,1^{+})$ possesses a large branching fraction for decays into the $\Upsilon B$ final state, whereas its decays to $\Upsilon B^*$ are suppressed. 
The $T_{b^{2}\bar{b}\bar{n}}(15125,1^{+})$ displays the inverted decay pattern.
From the perspective of radiative transitions, the radiative decay width for $T_{b^2\bar{b}\bar{n}}(15129,1^+)\to T_{b^2\bar{b}\bar{n}}(15057,1^+)$ amounts to 2.2$|$1.1 keV. 
In sharp contrast, the radiative transition $T_{b^2\bar{b}\bar{n}}(15125,1^+)\to T_{b^2\bar{b}\bar{n}}(15057,1^+)$ is strongly suppressed, with a partial width smaller than 0.05 keV. 
Thus, such radiative transition provides a viable criterion for differentiating \(T_{b^2\bar{b}\bar{n}}(15129,1^+)\) from \(T_{b^2\bar{b}\bar{n}}(15125,1^+)\).
Although we can theoretically distinguish them by branching ratios and radiative transition,
current experimental detectors still face substantial challenges in distinguishing nearly degenerate states with a mass difference ($\Delta M$) of 4 MeV. 
Moreover, $T_{b^{2}\bar{b}\bar{n}}(15057,1^{+})$ is the lowest-mass state among the three $J^{P}=1^{+}$ $bb\bar{b}\bar{n}$ states, with a total decay width of approximately 21 MeV.
Its corresponding relative partial decay width ratio is given by:
\begin{eqnarray}
\Gamma_{\Upsilon B^{*}}:\Gamma_{\Upsilon B}:\Gamma_{\eta_{b} B^{*}}=2:1:1.2.
\end{eqnarray}
From the above ratio, we notice that $\Upsilon B^{*}$ channel is the dominant decay mode for $T_{b^{2}\bar{b}\bar{n}}(15057,1^{+})$.

For the two $J^{P}=0^{+}$ $bb\bar{b}\bar{n}$ states, 
$T_{b^{2}\bar{b}\bar{n}}(15148,0^{+})$ is the highest-mass resonance in the $bb\bar{b}\bar{n}$ system.
Its total decay width is approximately 22 MeV, with the ratio of partial decay widths given by:
\begin{eqnarray}
\frac{\Gamma[T_{b^{2}\bar{b}\bar{n}}(15148,0^{+})\to \eta_{b} B]}{\Gamma[T_{b^{2}\bar{b}\bar{n}}(15148,0^{+})\to \Upsilon B^{*}]}=1.3.
\end{eqnarray}
Both the $\eta_{b} B$ and $\Upsilon B^{*}$ decay channels offer excellent prospects for experimental reconstruction of this resonance.
The other $J^{P}=0^{+}$ $bb\bar{b}\bar{n}$ state, $T_{b^2\bar{b}\bar{n}}(15052,0^+)$, also has a total decay width of 22 MeV.
Its relative partial decay width ratio is
\begin{eqnarray}
\frac{\Gamma[T_{b^{2}\bar{b}\bar{n}}(15052,0^{+})\to \Upsilon B^{*}]}{\Gamma[T_{b^{2}\bar{b}\bar{n}}(15052,0^{+})\to \eta_{b} B]}=1.2.
\end{eqnarray}
Thus, both $\eta_{b} B$ and $\Upsilon B^{*}$ decay channels are its main decay channels.
We therefore propose that experiments scan the invariant mass spectrum of $\eta_{b} B$ or $\Upsilon B^{*}$ in the mass range of 15.0-15.1 GeV, with special emphasis on the narrow peak near 15050 MeV.
If the peak position and width observed in the experiment are consistent with
theoretical predictions, such a peak would be identified as the signal for $T_{b^{2}\bar{b}\bar{n}}(15052,0^{+})$.
For the $bb\bar{b}\bar{s}$ system, 
an analogous analysis of decay properties can be carried out using Table \ref{QQQq3} and Fig. \ref{fig-bbbn}.

\section{Summary}\label{sec5}

Based on the nonrelativistic quark model, 
this study constructs an effective Hamiltonian including the confinement potential $V^{C}$ and color-spin hyperfine potential $V^{CS}$.
The four-body Schr\"{o}dinger equation is solved via the Gaussian expansion method, with the spatial wave function expanded in terms of Jacobi coordinates.
Furthermore, considering the effect of color-spin configuration mixing, 
this work systematically investigates the mass spectra, root-mean-square (RMS) radii, radiative decay, and rearrangement strong decay properties of the $cc\bar{c}\bar{n}$, $cc\bar{c}\bar{s}$, $bb\bar{b}\bar{n}$, and $bb\bar{b}\bar{s}$ tetraquark systems. 
Among these, the quark-interchange model is employed to calculate the widths of OZI-allowed two-body rearrangement strong decays, 
while tree-level quark-photon coupling is utilized to perform the theoretical calculation of radiative decay widths.
Finally, the complete structure and decay characteristics of the tetraquark states in each system are obtained.

It is found that, due to identical symmetry constraints, both the $cc\bar{c}\bar{q}$ and $bb\bar{b}\bar{q}$ systems possess two $J^{P}=0^{+}$, three $J^{P}=1^{+}$ and one $J^{P}=2^{+}$ ground states.
Further, the $cc\bar{c}\bar{q}$ and $bb\bar{b}\bar{q}$ systems exhibit analogous mass spectra because of the heavy quark symmetry.
The ground-state masses of these two systems fall in the ranges of 5.2–5.5 GeV and 15.0–15.3 GeV, respectively.
The larger constituent mass of the strange quark causes the states in the s-containing systems to be systematically heavier than their counterparts in the n-containing systems.
In contrast, the $V^{CS}$ interaction scales with $1/m_{i}m_{j}$, resulting in a relatively small magnitude of mass gaps between the corresponding states. 
In addition, the degree of color-spin configuration mixing in the $bb\bar{b}\bar{q}$ system is lower than that in the $cc\bar{c}\bar{q}$ system, and the mass gaps between the physical states after mixing are larger  compared to their pre-mixing counterparts. 
Moreover, all the quantum states investigated lie above the meson-meson decay thresholds, indicating that no stable bound state exists in these systems.

The RMS radius analysis shows that the interparticle separations in the $cc\bar{c}\bar{q}$ system range from 0.4 to 0.6 fm, with an intercluster distance of about 0.45 fm.
For the $bb\bar{b}\bar{q}$ system, the corresponding values are 0.2–0.5 fm and about 0.35 fm, respectively.
For both systems, the average intercluster separation is of the same order of magnitude as the separation between constituent (anti)quarks, exhibiting a prominent feature of strong spatial overlap. 
This characteristic supports the interpretation that these states correspond to compact tetraquark configurations.

Regarding decay properties, rearrangement strong decays constitute the dominant channels for all states, while radiative decay widths are generally negligible.
In particular, the radiative decay widths of states in the $bb\bar{b}\bar{q}$ system are generally below 2.5 keV, much smaller than those in the $cc\bar{c}\bar{q}$ system. 
The difference in electric charge between the $\bar{u}$ and $\bar{d}$ quarks leads to distinct disparities in the radiative decay widths of the $cc\bar{c}\bar{u}$/$cc\bar{c}\bar{d}$ and $bb\bar{b}\bar{u}$/$bb\bar{b}\bar{d}$ states. 
Nevertheless, due to the absolute dominance of rearrangement strong decays, the total decay widths of all these states are virtually identical.
Obvious distinctions exist in the decay patterns of states with different spin-parity quantum numbers: the $J^{P}=2^{+}$ states, constrained by angular momentum conservation, decay mainly into vector meson–vector meson final states via an $S$-wave, while their $D$-wave decay channels are suppressed; 
additionally, they cannot undergo radiative transitions to the $J^{P}=0^{+}$ states. 
Some partner states with identical quantum numbers exhibit extremely small mass differences.
For instance, the mass gap between the two highest-mass $J^{P}=1^{+}$ states in the $cc\bar{c}\bar{s}$ ($bb\bar{b}\bar{n}$) system is merely 10 (4) MeV. 
Therefore, the marked differences in their radiative transitions and decay branching ratios provide key experimental signatures for distinguishing such nearly degenerate states.

This study also identifies a number of narrow states with distinct characteristic features;
the intrinsic formation mechanism for such states is that, despite their larger decay phase space, the Feynman amplitudes $\mathcal{M}(A\to BC)$ derived from the four distinct quark-interchange diagrams possess opposite signs, 
leading to substantial mutual cancellation of their respective contributions.
Notable examples are $T_{c^{2}\bar{c}\bar{s}}(5360,0^{+})$ and $T_{b^{2}\bar{b}\bar{n}}(15148,2^{+})$, both of which are narrow resonance states with a total decay width of less than 20 MeV. 
For the former state, the $J/\psi D^{*}_{s}$ and $\eta_{c} D_{s}$ channels are decay modes of equal strength, while the $\Upsilon B^{*}$ channel is the dominant decay mode for the latter.
Based on the decay characteristics of each state, 
targeted experimental search proposals are put forward in this work: 
the invariant mass spectra of $J/\psi D^{*}_{s}$ and $\eta_{c} D_{s}$ should be scanned in the mass range of 5.3–5.4 GeV to search for $T_{c^{2}\bar{c}\bar{s}}(5360,0^{+})$, and the invariant mass spectrum of $\Upsilon B^{*}$ should be scanned in the 15.1–15.2 GeV range to look for $T_{b^{2}\bar{b}\bar{n}}(15148,2^{+})$.
Signal confirmation requires a consistent match between the observed peak position, decay width, and decay branching ratio.

In summary, 
our results comprehensively reveal the mass spectra, internal structures, and decay characteristics of $cc\bar{c}\bar{q}$ and $bb\bar{b}\bar{q}$ tetraquark states, 
and provide detailed theoretical predictions as well as explicit experimental search channels for the experimental exploration and identification of such states. 
The framework employed in this work, which combines the Gaussian expansion method with the quark-interchange model, also provides an important theoretical reference for studies of the spectroscopic and decay properties of other exotic hadrons. 
These findings can offer valuable guidance for both theorists and experimentalists in the field of heavy-flavor hadron physics,
whereas more in-depth dynamical investigations of the $cc\bar{c}\bar{q}$ and $bb\bar{b}\bar{q}$ tetraquark states remain imperative.
Finally, we hope that the present study can motivate the LHCb, Belle II, BESIII and other relevant experimental collaborations to perform dedicated searches for the $cc\bar{c}\bar{q}$ and $bb\bar{b}\bar{q}$ tetraquark states. 
Conducting more targeted experimental measurements will not only test our theoretical predictions but also further deepen the understanding of the internal interaction mechanisms of such tetraquark states, and provide insightful perspectives for future theoretical exploration of heavy-flavored exotic hadron physics.
\\

\section*{Acknowledgements}
This work is supported by the National Nature Science Foundation of China under Grant No.12447172, 12447155, and 12405098, by the Postdoctoral Fellowship Program of CPSF under Grant No.GZC20240877, 2025M773368, and GZC20240056 and by Shuimu Tsinghua Scholar Program of Tsinghua University under Grant No.2024SM119.

\bibliographystyle{UserDefined}
\bibliography{References}

@article{Berwein:2024ztx,
    author = "Berwein, Matthias and Brambilla, Nora and Mohapatra, Abhishek and Vairo, Antonio",
    title = "{Hybrids, tetraquarks, pentaquarks, doubly heavy baryons, and quarkonia in Born-Oppenheimer effective theory}",
    eprint = "2408.04719",
    archivePrefix = "arXiv",
    primaryClass = "hep-ph",
    reportNumber = "TUM-EFT 185/23",
    doi = "10.1103/PhysRevD.110.094040",
    journal = "Phys. Rev. D",
    volume = "110",
    number = "9",
    pages = "094040",
    year = "2024"
}

@article{Wu:2024euj,
    author = "Wu, Wei-Lin and Chen, Yan-Ke and Meng, Lu and Zhu, Shi-Lin",
    title = "{Benchmark calculations of fully heavy compact and molecular tetraquark states}",
    eprint = "2401.14899",
    archivePrefix = "arXiv",
    primaryClass = "hep-ph",
    doi = "10.1103/PhysRevD.109.054034",
    journal = "Phys. Rev. D",
    volume = "109",
    number = "5",
    pages = "054034",
    year = "2024"
}

@article{CMS:2025fpt,
    author = "Hayrapetyan, Aram and others",
    collaboration = "CMS",
    title = "{Determination of the spin and parity of all-charm tetraquarks}",
    eprint = "2506.07944",
    archivePrefix = "arXiv",
    primaryClass = "hep-ex",
    reportNumber = "CMS-BPH-24-002, CERN-EP-2025-118",
    doi = "10.1038/s41586-025-09711-7",
    journal = "Nature",
    volume = "648",
    number = "8092",
    pages = "58--63",
    year = "2025"
}

@article{BESIII:2023wsc,
    author = "Ablikim, M. and others",
    collaboration = "BESIII",
    title = "{Precise Measurement of the e+e-{\textrightarrow}Ds*+Ds*- Cross Sections at Center-of-Mass Energies from Threshold to 4.95~GeV}",
    eprint = "2305.10789",
    archivePrefix = "arXiv",
    primaryClass = "hep-ex",
    doi = "10.1103/PhysRevLett.131.151903",
    journal = "Phys. Rev. Lett.",
    volume = "131",
    number = "15",
    pages = "151903",
    year = "2023"
}

@article{Galkin:2025ubt,
    author = "Galkin, V. O. and Savchenko, E. M.",
    title = "{Masses of Ground States of Triply Heavy Tetraquarks}",
    doi = "10.1134/S1063779624701594",
    journal = "Phys. Part. Nucl.",
    volume = "56",
    number = "2",
    pages = "330--334",
    year = "2025"
}

@article{Zhang:2024jvv,
    author = "Zhang, Wen-Shuai and Tang, Liang",
    title = "{Investigating triply heavy tetraquark states through QCD sum rules}",
    eprint = "2412.11531",
    archivePrefix = "arXiv",
    primaryClass = "hep-ph",
    doi = "10.1016/j.nuclphysa.2025.123227",
    journal = "Nucl. Phys. A",
    volume = "1064",
    pages = "123227",
    year = "2025"
}

@article{Chen:2016ont,
    author = "Chen, Kan and Liu, Xiang and Wu, Jing and Liu, Yan-Rui and Zhu, Shi-Lin",
    title = "{Triply heavy tetraquark states with the $QQ\bar{Q}\bar{q}$ configuration}",
    eprint = "1609.06117",
    archivePrefix = "arXiv",
    primaryClass = "hep-ph",
    doi = "10.1140/epja/i2017-12199-3",
    journal = "Eur. Phys. J. A",
    volume = "53",
    number = "1",
    pages = "5",
    year = "2017"
}

@article{Liu:2022jdl,
    author = "Liu, Xuejie and Tan, Yue and Chen, Dianyong and Huang, Hongxia and Ping, Jialun",
    title = "{Possible triply heavy tetraquark states in a chiral quark model}",
    eprint = "2205.08281",
    archivePrefix = "arXiv",
    primaryClass = "hep-ph",
    doi = "10.1103/PhysRevD.107.054019",
    journal = "Phys. Rev. D",
    volume = "107",
    number = "5",
    pages = "054019",
    year = "2023"
}

@article{Weng:2021ngd,
    author = "Weng, Xin-Zhen and Deng, Wei-Zhen and Zhu, Shi-Lin",
    title = "{Triply heavy tetraquark states}",
    eprint = "2109.05243",
    archivePrefix = "arXiv",
    primaryClass = "hep-ph",
    doi = "10.1103/PhysRevD.105.034026",
    journal = "Phys. Rev. D",
    volume = "105",
    number = "3",
    pages = "034026",
    year = "2022"
}

@article{Silvestre-Brac:1993zem,
    author = "Silvestre-Brac, B. and Semay, C.",
    title = "{Systematics of L = 0 $q^2 \bar{q}^2$ systems}",
    doi = "10.1007/BF01565058",
    journal = "Z. Phys. C",
    volume = "57",
    pages = "273--282",
    year = "1993"
}

@article{Junnarkar:2018twb,
    author = "Junnarkar, Parikshit and Mathur, Nilmani and Padmanath, M.",
    title = "{Study of doubly heavy tetraquarks in Lattice QCD}",
    eprint = "1810.12285",
    archivePrefix = "arXiv",
    primaryClass = "hep-lat",
    reportNumber = "TIFR/TH/18-43",
    doi = "10.1103/PhysRevD.99.034507",
    journal = "Phys. Rev. D",
    volume = "99",
    number = "3",
    pages = "034507",
    year = "2019"
}

@article{Li:1994cy,
    author = "Li, Zhen-Ping",
    title = "{The Threshold pion photoproduction of nucleons in the chiral quark model}",
    eprint = "hep-ph/9404269",
    archivePrefix = "arXiv",
    doi = "10.1103/PhysRevD.50.5639",
    journal = "Phys. Rev. D",
    volume = "50",
    pages = "5639--5646",
    year = "1994"
}

@article{Davila-Rivera:2025exk,
    author = "D{\'a}vila-Rivera, A. and Garc{\'\i}a-Tecocoatzi, H. and Ramirez-Morales, A. and Rivero-Acosta, Ailier and Santopinto, E. and Vaquera-Araujo, Carlos Alberto",
    title = "{Radiative decays of the $\Sigma_c$, $\Xi'_c$ and $\Omega_c$ charmed baryons}",
    eprint = "2512.03008",
    archivePrefix = "arXiv",
    primaryClass = "hep-ph",
    month = "12",
    year = "2025"
}

@article{Peng:2024pyl,
    author = "Peng, Yu-Xin and Luo, Si-Qiang and Liu, Xiang",
    title = "{Refining radiative decay studies in singly heavy baryons}",
    eprint = "2405.12812",
    archivePrefix = "arXiv",
    primaryClass = "hep-ph",
    doi = "10.1103/PhysRevD.110.074034",
    journal = "Phys. Rev. D",
    volume = "110",
    number = "7",
    pages = "074034",
    year = "2024"
}

@article{Wang:2023ael,
    author = "Wang, Fu-Lai and Liu, Xiang",
    title = "{Surveying the mass spectra and the electromagnetic properties of the $\Xi_{c}^{(\prime,*)}D^{(*)}$ molecular pentaquarks}",
    eprint = "2311.13968",
    archivePrefix = "arXiv",
    primaryClass = "hep-ph",
    doi = "10.1103/PhysRevD.109.014043",
    journal = "Phys. Rev. D",
    volume = "109",
    number = "1",
    pages = "014043",
    year = "2024"
}

@article{Zhang:2025ame,
    author = "Zhang, Chen-Ke and Wang, Fu-Lai and Luo, Si-Qiang and Liu, Xiang",
    title = "{Electromagnetic properties of possible triple-charm molecular hexaquarks}",
    eprint = "2505.10318",
    archivePrefix = "arXiv",
    primaryClass = "hep-ph",
    doi = "10.1140/epjc/s10052-025-14317-4",
    journal = "Eur. Phys. J. C",
    volume = "85",
    number = "5",
    pages = "582",
    year = "2025"
}

@article{Sheng:2024hkf,
    author = "Sheng, Li-Cheng and Huo, Jin-Yu and Chen, Rui and Wang, Fu-Lai and Liu, Xiang",
    title = "{Exploring the mass spectrum and electromagnetic property of the $\Xi_{cc}K^{(*)}$ and $\Xi_{cc}\bar{K}^{(*)}$ molecules}",
    eprint = "2406.16115",
    archivePrefix = "arXiv",
    primaryClass = "hep-ph",
    doi = "10.1103/PhysRevD.110.054044",
    journal = "Phys. Rev. D",
    volume = "110",
    number = "5",
    pages = "054044",
    year = "2024"
}

@article{Zhou:2025fpp,
    author = "Zhou, Hao and Luo, Si-Qiang and Liu, Xiang",
    title = "{Triply heavy baryon spectroscopy revisited}",
    eprint = "2507.10243",
    archivePrefix = "arXiv",
    primaryClass = "hep-ph",
    doi = "10.1103/jhr1-ccsw",
    journal = "Phys. Rev. D",
    volume = "112",
    number = "7",
    pages = "074007",
    year = "2025"
}

@article{Zhao:2002id,
    author = "Zhao, Qiang and Al-Khalili, J. S. and Li, Z. P. and Workman, R. L.",
    title = "{Pion photoproduction on the nucleon in the quark model}",
    eprint = "nucl-th/0202067",
    archivePrefix = "arXiv",
    doi = "10.1103/PhysRevC.65.065204",
    journal = "Phys. Rev. C",
    volume = "65",
    pages = "065204",
    year = "2002"
}

@article{Lu:2017meb,
    author = {L{\"u}, Qi-Fang and Wang, Kai-Lei and Xiao, Li-Ye and Zhong, Xian-Hui},
    title = "{Mass spectra and radiative transitions of doubly heavy baryons in a relativized quark model}",
    eprint = "1708.04468",
    archivePrefix = "arXiv",
    primaryClass = "hep-ph",
    doi = "10.1103/PhysRevD.96.114006",
    journal = "Phys. Rev. D",
    volume = "96",
    number = "11",
    pages = "114006",
    year = "2017"
}

@article{Wang:2017kfr,
    author = "Wang, Kai-Lei and Yao, Ya-Xiong and Zhong, Xian-Hui and Zhao, Qiang",
    title = "{Strong and radiative decays of the low-lying $S$- and $P$-wave singly heavy baryons}",
    eprint = "1709.04268",
    archivePrefix = "arXiv",
    primaryClass = "hep-ph",
    doi = "10.1103/PhysRevD.96.116016",
    journal = "Phys. Rev. D",
    volume = "96",
    number = "11",
    pages = "116016",
    year = "2017"
}

@article{Yao:2018jmc,
    author = "Yao, Ya-Xiong and Wang, Kai-Lei and Zhong, Xian-Hui",
    title = "{Strong and radiative decays of the low-lying $D$-wave singly heavy baryons}",
    eprint = "1803.00364",
    archivePrefix = "arXiv",
    primaryClass = "hep-ph",
    doi = "10.1103/PhysRevD.98.076015",
    journal = "Phys. Rev. D",
    volume = "98",
    number = "7",
    pages = "076015",
    year = "2018"
}

@article{Xiao:2017udy,
    author = "Xiao, Li-Ye and Wang, Kai-Lei and Lu, Qi-fang and Zhong, Xian-Hui and Zhu, Shi-Lin",
    title = "{Strong and radiative decays of the doubly charmed baryons}",
    eprint = "1708.04384",
    archivePrefix = "arXiv",
    primaryClass = "hep-ph",
    doi = "10.1103/PhysRevD.96.094005",
    journal = "Phys. Rev. D",
    volume = "96",
    number = "9",
    pages = "094005",
    year = "2017"
}

@article{Li:1997gd,
    author = "Li, Zhen-ping and Ye, Hong-xing and Lu, Ming-hui",
    title = "{An Unified approach to pseudoscalar meson photoproductions off nucleons in the quark model}",
    eprint = "nucl-th/9706010",
    archivePrefix = "arXiv",
    doi = "10.1103/PhysRevC.56.1099",
    journal = "Phys. Rev. C",
    volume = "56",
    pages = "1099--1113",
    year = "1997"
}

@article{Deng:2016ktl,
    author = "Deng, Wei-Jun and Liu, Hui and Gui, Long-Cheng and Zhong, Xian-Hui",
    title = "{Spectrum and electromagnetic transitions of bottomonium}",
    eprint = "1607.04696",
    archivePrefix = "arXiv",
    primaryClass = "hep-ph",
    doi = "10.1103/PhysRevD.95.074002",
    journal = "Phys. Rev. D",
    volume = "95",
    number = "7",
    pages = "074002",
    year = "2017"
}

@article{Deng:2016stx,
    author = "Deng, Wei-Jun and Liu, Hui and Gui, Long-Cheng and Zhong, Xian-Hui",
    title = "{Charmonium spectrum and their electromagnetic transitions with higher multipole contributions}",
    eprint = "1608.00287",
    archivePrefix = "arXiv",
    primaryClass = "hep-ph",
    doi = "10.1103/PhysRevD.95.034026",
    journal = "Phys. Rev. D",
    volume = "95",
    number = "3",
    pages = "034026",
    year = "2017"
}

@article{Li:2025wod,
    author = "Li, Na and Xing, Ye and Shi, Jing-Rui",
    title = "{Analysis of triply charmed four-quark state in the effective Lagrangian approach}",
    eprint = "2503.03555",
    archivePrefix = "arXiv",
    primaryClass = "hep-ph",
    month = "3",
    year = "2025"
}

@article{Li:2025fmf,
    author = "Li, Shi-Yuan and Liu, Yan-Rui and Man, Zi-Long and Shu, Cheng-Rui and Si, Zong-Guo and Wu, Jing",
    title = "{Triply Heavy Tetraquark States in a Mass-Splitting Model}",
    eprint = "2501.16105",
    archivePrefix = "arXiv",
    primaryClass = "hep-ph",
    doi = "10.3390/sym17020170",
    journal = "Symmetry",
    volume = "17",
    number = "2",
    pages = "170",
    year = "2025"
}

@article{Cheng:2020nho,
    author = "Cheng, Jian-Bo and Li, Shi-Yuan and Liu, Yan-Rui and Liu, Yu-Nan and Si, Zong-Guo and Yao, Tao",
    title = "{Spectrum and rearrangement decays of tetraquark states with four different flavors}",
    eprint = "2001.05287",
    archivePrefix = "arXiv",
    primaryClass = "hep-ph",
    doi = "10.1103/PhysRevD.101.114017",
    journal = "Phys. Rev. D",
    volume = "101",
    number = "11",
    pages = "114017",
    year = "2020"
}

@article{Yang:2025wqo,
    author = "Yang, Hui-Min and Ma, Yao and Wu, Wei-Lin and Zhu, Shi-Lin",
    title = "{Triply heavy tetraquark states with different flavors}",
    eprint = "2502.10798",
    archivePrefix = "arXiv",
    primaryClass = "hep-ph",
    doi = "10.1103/PhysRevD.111.074040",
    journal = "Phys. Rev. D",
    volume = "111",
    number = "7",
    pages = "074040",
    year = "2025"
}

@article{Xu:2025zna,
    author = "Xu, Yong-Jiang and Tian, Luo-Geng and Li, Yuan and Wu, Jin-Yun and Liu, Yong-Lu and Huang, Ming-Qiu",
    title = "{Scalar Triple-Heavy Tetraquark States With Quark Content $cc\bar{c}\bar{s}$}",
    eprint = "2506.14527",
    archivePrefix = "arXiv",
    primaryClass = "hep-ph",
    month = "6",
    year = "2025"
}

@article{Xing:2019wil,
    author = "Xing, Ye",
    title = "{Weak decays of triply heavy tetraquarks ${b{\bar{c}}}{b{\bar{q}}}$}",
    eprint = "1910.11593",
    archivePrefix = "arXiv",
    primaryClass = "hep-ph",
    doi = "10.1140/epjc/s10052-020-7625-3",
    journal = "Eur. Phys. J. C",
    volume = "80",
    number = "1",
    pages = "57",
    year = "2020"
}

@article{Jiang:2017tdc,
    author = "Jiang, Jin-Feng and Chen, Wei and Zhu, Shi-Lin",
    title = "{Triply heavy $QQ\bar Q\bar q$ tetraquark states}",
    eprint = "1708.00142",
    archivePrefix = "arXiv",
    primaryClass = "hep-ph",
    doi = "10.1103/PhysRevD.96.094022",
    journal = "Phys. Rev. D",
    volume = "96",
    number = "9",
    pages = "094022",
    year = "2017"
}

@article{Zhu:2023lbx,
    author = "Zhu, Zhen-Hui and Zhang, Wen-Xuan and Jia, Duojie",
    title = "{Triply heavy tetraquark states: masses and other properties}",
    eprint = "2312.01908",
    archivePrefix = "arXiv",
    primaryClass = "hep-ph",
    doi = "10.1140/epjc/s10052-024-12700-1",
    journal = "Eur. Phys. J. C",
    volume = "84",
    number = "4",
    pages = "344",
    year = "2024"
}

@article{Mutuk:2023yev,
    author = "Mutuk, Halil",
    title = "{Flavor exotic triply-heavy tetraquark states in AdS/QCD potential}",
    eprint = "2305.03358",
    archivePrefix = "arXiv",
    primaryClass = "hep-ph",
    doi = "10.1140/epjc/s10052-023-11526-7",
    journal = "Eur. Phys. J. C",
    volume = "83",
    number = "5",
    pages = "358",
    year = "2023"
}

@article{Liu:2024ziu,
    author = "Liu, Ming-Zhu and Ling, Xi-Zhe and Geng, Li-Sheng",
    title = "{Productions of X(3872)/Zc(3900) and X2(4013)/Zc(4020) in Y(4220) and Y(4360) decays}",
    eprint = "2404.07681",
    archivePrefix = "arXiv",
    primaryClass = "hep-ph",
    doi = "10.1103/PhysRevD.110.054035",
    journal = "Phys. Rev. D",
    volume = "110",
    number = "5",
    pages = "054035",
    year = "2024"
}

@article{Hudspith:2020tdf,
    author = "Hudspith, R. J. and Colquhoun, B. and Francis, A. and Lewis, R. and Maltman, K.",
    title = "{A lattice investigation of exotic tetraquark channels}",
    eprint = "2006.14294",
    archivePrefix = "arXiv",
    primaryClass = "hep-lat",
    doi = "10.1103/PhysRevD.102.114506",
    journal = "Phys. Rev. D",
    volume = "102",
    pages = "114506",
    year = "2020"
}

@article{Yang:2024nyc,
    author = "Yang, Gang and Ping, Jialun and Segovia, Jorge",
    title = "{Triply charm and bottom tetraquarks in a constituent quark model}",
    eprint = "2407.14548",
    archivePrefix = "arXiv",
    primaryClass = "hep-ph",
    doi = "10.1103/PhysRevD.110.054036",
    journal = "Phys. Rev. D",
    volume = "110",
    number = "5",
    pages = "054036",
    year = "2024"
}

@article{Lu:2021kut,
    author = {L{\"u}, Qi-Fang and Chen, Dian-Yong and Dong, Yu-Bing and Santopinto, Elena},
    title = "{Triply-heavy tetraquarks in an extended relativized quark model}",
    eprint = "2107.13930",
    archivePrefix = "arXiv",
    primaryClass = "hep-ph",
    doi = "10.1103/PhysRevD.104.054026",
    journal = "Phys. Rev. D",
    volume = "104",
    number = "5",
    pages = "054026",
    year = "2021"
}

@article{LHCb:2020bls,
    author = "Aaij, Roel and others",
    collaboration = "LHCb",
    title = "{A model-independent study of resonant structure in $B^+\to D^+D^-K^+$ decays}",
    eprint = "2009.00025",
    archivePrefix = "arXiv",
    primaryClass = "hep-ex",
    reportNumber = "LHCb-PAPER-2020-024, CERN-EP-2020-158",
    doi = "10.1103/PhysRevLett.125.242001",
    journal = "Phys. Rev. Lett.",
    volume = "125",
    pages = "242001",
    year = "2020"
}

@article{LHCb:2020pxc,
    author = "Aaij, Roel and others",
    collaboration = "LHCb",
    title = "{Amplitude analysis of the $B^+\to D^+D^-K^+$ decay}",
    eprint = "2009.00026",
    archivePrefix = "arXiv",
    primaryClass = "hep-ex",
    reportNumber = "LHCb-PAPER-2020-025, CERN-EP-2020-159",
    doi = "10.1103/PhysRevD.102.112003",
    journal = "Phys. Rev. D",
    volume = "102",
    pages = "112003",
    year = "2020"
}

@article{LHCb:2020bwg,
    author = "Aaij, Roel and others",
    collaboration = "LHCb",
    title = "{Observation of structure in the $J /\psi$ -pair mass spectrum}",
    eprint = "2006.16957",
    archivePrefix = "arXiv",
    primaryClass = "hep-ex",
    reportNumber = "CERN-EP-2020-115, LHCb-PAPER-2020-011",
    doi = "10.1016/j.scib.2020.08.032",
    journal = "Sci. Bull.",
    volume = "65",
    number = "23",
    pages = "1983--1993",
    year = "2020"
}

@article{CMS:2023owd,
    author = "Hayrapetyan, Aram and others",
    collaboration = "CMS",
    title = "{New Structures in the J/{\ensuremath{\psi}}J/{\ensuremath{\psi}} Mass Spectrum in Proton-Proton Collisions at s=13{\,}{\,}TeV}",
    eprint = "2306.07164",
    archivePrefix = "arXiv",
    primaryClass = "hep-ex",
    reportNumber = "CMS-BPH-21-003, CERN-EP-2023-109",
    doi = "10.1103/PhysRevLett.132.111901",
    journal = "Phys. Rev. Lett.",
    volume = "132",
    number = "11",
    pages = "111901",
    year = "2024"
}

@article{ATLAS:2023bft,
    author = "Aad, Georges and others",
    collaboration = "ATLAS",
    title = "{Observation of an Excess of Dicharmonium Events in the Four-Muon Final State with the ATLAS Detector}",
    eprint = "2304.08962",
    archivePrefix = "arXiv",
    primaryClass = "hep-ex",
    reportNumber = "CERN-EP-2023-035",
    doi = "10.1103/PhysRevLett.131.151902",
    journal = "Phys. Rev. Lett.",
    volume = "131",
    number = "15",
    pages = "151902",
    year = "2023"
}

@article{Huang:2020dci,
    author = "Huang, Guojun and Zhao, Jiaxing and Zhuang, Pengfei",
    title = "{Pair structure of heavy tetraquark systems}",
    eprint = "2012.14845",
    archivePrefix = "arXiv",
    primaryClass = "hep-ph",
    doi = "10.1103/PhysRevD.103.054014",
    journal = "Phys. Rev. D",
    volume = "103",
    number = "5",
    pages = "054014",
    year = "2021"
}

@article{Yang:2020wkh,
    author = "Yang, Bo-Cheng and Tang, Liang and Qiao, Cong-Feng",
    title = "{Scalar fully-heavy tetraquark states $QQ^\prime {\bar{Q}} \bar{Q^\prime }$ in QCD sum rules}",
    eprint = "2012.04463",
    archivePrefix = "arXiv",
    primaryClass = "hep-ph",
    doi = "10.1140/epjc/s10052-021-09096-7",
    journal = "Eur. Phys. J. C",
    volume = "81",
    number = "4",
    pages = "324",
    year = "2021"
}

@article{Lu:2020cns,
    author = {L{\"u}, Qi-Fang and Chen, Dian-Yong and Dong, Yu-Bing},
    title = "{Masses of fully heavy tetraquarks $QQ {\bar{Q}} {\bar{Q}}$ in an extended relativized quark model}",
    eprint = "2006.14445",
    archivePrefix = "arXiv",
    primaryClass = "hep-ph",
    doi = "10.1140/epjc/s10052-020-08454-1",
    journal = "Eur. Phys. J. C",
    volume = "80",
    number = "9",
    pages = "871",
    year = "2020"
}

@article{Albuquerque:2020hio,
    author = "Albuquerque, R. M. and Narison, S. and Rabemananjara, A. and Rabetiarivony, D. and Randriamanatrika, G.",
    title = "{Doubly-hidden scalar heavy molecules and tetraquarks states from QCD at NLO}",
    eprint = "2008.01569",
    archivePrefix = "arXiv",
    primaryClass = "hep-ph",
    doi = "10.1103/PhysRevD.102.094001",
    journal = "Phys. Rev. D",
    volume = "102",
    number = "9",
    pages = "094001",
    year = "2020"
}

@article{LHCb:2022lzp,
    author = "Aaij, R. and others",
    collaboration = "LHCb",
    title = "{Amplitude analysis of $B^{0}\to \bar{D}^{0}D_{s}^{+}\pi^{-}$ and $B^{+}\to D^{-}D_{s}^{+}\pi^{+}$ decays}",
    eprint = "2212.02717",
    archivePrefix = "arXiv",
    primaryClass = "hep-ex",
    reportNumber = "LHCb-PAPER-2022-027, CERN-EP-2022-246",
    doi = "10.1103/PhysRevD.108.012017",
    journal = "Phys. Rev. D",
    volume = "108",
    number = "1",
    pages = "012017",
    year = "2023"
}

@article{Cheng:2003kg,
    author = "Cheng, Hai-Yang and Hou, Wei-Shu",
    title = "{B decays as spectroscope for charmed four quark states}",
    eprint = "hep-ph/0305038",
    archivePrefix = "arXiv",
    doi = "10.1016/S0370-2693(03)00834-7",
    journal = "Phys. Lett. B",
    volume = "566",
    pages = "193--200",
    year = "2003"
}

@article{Kim:2005gt,
    author = "Kim, Hungchong and Oh, Yongseok",
    title = "{D(s)(2317) as a four-quark state in QCD sum rules}",
    eprint = "hep-ph/0508251",
    archivePrefix = "arXiv",
    doi = "10.1103/PhysRevD.72.074012",
    journal = "Phys. Rev. D",
    volume = "72",
    pages = "074012",
    year = "2005"
}

@article{Guo:2006rp,
    author = "Guo, Feng-Kun and Shen, Peng-Nian and Chiang, Huan-Ching",
    title = "{Dynamically generated 1+ heavy mesons}",
    eprint = "hep-ph/0610008",
    archivePrefix = "arXiv",
    doi = "10.1016/j.physletb.2007.01.050",
    journal = "Phys. Lett. B",
    volume = "647",
    pages = "133--139",
    year = "2007"
}

@article{Navarra:2015iea,
    author = "Navarra, Fernando S. and Nielsen, Marina and Oset, Eulogio and Sekihara, Takayasu",
    title = "{Testing the molecular nature of $D_{s0}^*(2317)$ and $D_0^*(2400)$ in semileptonic $B_s$ and $B$ decays}",
    eprint = "1501.03422",
    archivePrefix = "arXiv",
    primaryClass = "hep-ph",
    doi = "10.1103/PhysRevD.92.014031",
    journal = "Phys. Rev. D",
    volume = "92",
    number = "1",
    pages = "014031",
    year = "2015"
}

@article{Chen:2004dy,
    author = "Chen, Yu-Qi and Li, Xue-Qian",
    title = "{A Comprehensive four-quark interpretation of $D_{s}(2317)$, $D_{s}(2457)$ and $D_{s}(2632)$}",
    eprint = "hep-ph/0407062",
    archivePrefix = "arXiv",
    doi = "10.1103/PhysRevLett.93.232001",
    journal = "Phys. Rev. Lett.",
    volume = "93",
    pages = "232001",
    year = "2004"
}

@article{Barnes:2003dj,
    author = "Barnes, T. and Close, F. E. and Lipkin, H. J.",
    title = "{Implications of a DK molecule at 2.32-GeV}",
    eprint = "hep-ph/0305025",
    archivePrefix = "arXiv",
    doi = "10.1103/PhysRevD.68.054006",
    journal = "Phys. Rev. D",
    volume = "68",
    pages = "054006",
    year = "2003"
}

@article{BaBar:2003oey,
    author = "Aubert, B. and others",
    collaboration = "BaBar",
    title = "{Observation of a narrow meson decaying to $D_s^+ \pi^0$ at a mass of 2.32-GeV/c$^2$}",
    eprint = "hep-ex/0304021",
    archivePrefix = "arXiv",
    reportNumber = "SLAC-PUB-9711, BABAR-PUB-03-011",
    doi = "10.1103/PhysRevLett.90.242001",
    journal = "Phys. Rev. Lett.",
    volume = "90",
    pages = "242001",
    year = "2003"
}

@article{CLEO:2003ggt,
    author = "Besson, D. and others",
    collaboration = "CLEO",
    title = "{Observation of a narrow resonance of mass 2.46 $\text{GeV}/c^{2}$ decaying to $D_{s}^{*+}\pi^0$ and confirmation of the $D^{*}_{sJ}(2317)$ state}",
    eprint = "hep-ex/0305100",
    archivePrefix = "arXiv",
    reportNumber = "CLNS-03-1826, CLEO-03-09, CLNS03-1826",
    doi = "10.1103/PhysRevD.68.032002",
    journal = "Phys. Rev. D",
    volume = "68",
    pages = "032002",
    year = "2003",
    note = "[Erratum: Phys.Rev.D 75, 119908 (2007)]"
}

@article{Dubnicka:2011mm,
    author = "Dubnicka, Stanislav and Dubnickova, Anna Z. and Ivanov, Mikhail A. and Koerner, Juergen G. and Santorelli, Pietro and Saidullaeva, Gozyal G.",
    title = "{One-photon decay of the tetraquark state $X(3872) \to \gamma + J/\psi$ in a relativistic constituent quark model with infrared confinement}",
    eprint = "1104.3974",
    archivePrefix = "arXiv",
    primaryClass = "hep-ph",
    reportNumber = "DSF-5-2011, MZ-TH-11-09",
    doi = "10.1103/PhysRevD.84.014006",
    journal = "Phys. Rev. D",
    volume = "84",
    pages = "014006",
    year = "2011"
}

@article{Wu:2016gas,
    author = "Wu, Jing and Liu, Yan-Rui and Chen, Kan and Liu, Xiang and Zhu, Shi-Lin",
    title = "{$X(4140)$, $X(4270)$, $X(4500)$ and $X(4700)$ and their $cs\bar{c}\bar{s}$ tetraquark partners}",
    eprint = "1608.07900",
    archivePrefix = "arXiv",
    primaryClass = "hep-ph",
    doi = "10.1103/PhysRevD.94.094031",
    journal = "Phys. Rev. D",
    volume = "94",
    number = "9",
    pages = "094031",
    year = "2016"
}

@article{Wang:2023ivd,
    author = "Wang, Zhi-Peng and Wang, Fu-Lai and Wang, Guang-Juan and Liu, Xiang",
    title = "{Probing exotic $J^{PC}$ resonances from deeply bound charmoniumlike molecules: Insights for identifying exotic hadrons}",
    eprint = "2312.03512",
    archivePrefix = "arXiv",
    primaryClass = "hep-ph",
    reportNumber = "KEK-TH-2580",
    doi = "10.1103/PhysRevD.110.L051501",
    journal = "Phys. Rev. D",
    volume = "110",
    number = "5",
    pages = "L051501",
    year = "2024"
}

@article{Liu:2013waa,
    author = "Liu, Xiang",
    title = "{An overview of $XYZ$ new particles}",
    eprint = "1312.7408",
    archivePrefix = "arXiv",
    primaryClass = "hep-ph",
    doi = "10.1007/s11434-014-0407-2",
    journal = "Chin. Sci. Bull.",
    volume = "59",
    pages = "3815--3830",
    year = "2014"
}

@article{Godfrey:2008nc,
    author = "Godfrey, Stephen and Olsen, Stephen L.",
    title = "{The Exotic XYZ Charmonium-like Mesons}",
    eprint = "0801.3867",
    archivePrefix = "arXiv",
    primaryClass = "hep-ph",
    doi = "10.1146/annurev.nucl.58.110707.171145",
    journal = "Ann. Rev. Nucl. Part. Sci.",
    volume = "58",
    pages = "51--73",
    year = "2008"
}

@article{Wang:2021aql,
    author = "Wang, Fu-Lai and Yang, Xin-Dian and Chen, Rui and Liu, Xiang",
    title = "{Correlation of the hidden-charm molecular tetraquarks and the charmoniumlike structures existing in the $B\to XYZ+K$ process}",
    eprint = "2103.04698",
    archivePrefix = "arXiv",
    primaryClass = "hep-ph",
    doi = "10.1103/PhysRevD.104.094010",
    journal = "Phys. Rev. D",
    volume = "104",
    number = "9",
    pages = "094010",
    year = "2021"
}

@article{Dubnicka:2010kz,
    author = "Dubnicka, Stanislav and Dubnickova, Anna Z. and Ivanov, Mikhail A. and Korner, Juergen G.",
    title = "{Quark model description of the tetraquark state X(3872) in a relativistic constituent quark model with infrared confinement}",
    eprint = "1004.1291",
    archivePrefix = "arXiv",
    primaryClass = "hep-ph",
    doi = "10.1103/PhysRevD.81.114007",
    journal = "Phys. Rev. D",
    volume = "81",
    pages = "114007",
    year = "2010"
}

@article{Wang:2013kva,
    author = "Wang, P. and Wang, X. G.",
    title = "{Study on X(3872) from effective field theory with pion exchange interaction}",
    eprint = "1304.0846",
    archivePrefix = "arXiv",
    primaryClass = "hep-ph",
    doi = "10.1103/PhysRevLett.111.042002",
    journal = "Phys. Rev. Lett.",
    volume = "111",
    number = "4",
    pages = "042002",
    year = "2013"
}

@article{BESIII:2013mhi,
    author = "Ablikim, M. and others",
    collaboration = "BESIII",
    title = "{Observation of a charged charmoniumlike structure in $e^+e^- \to (D^{*} \bar{D}^{*})^{\pm} \pi^\mp$ at $\sqrt{s}=4.26$GeV}",
    eprint = "1308.2760",
    archivePrefix = "arXiv",
    primaryClass = "hep-ex",
    doi = "10.1103/PhysRevLett.112.132001",
    journal = "Phys. Rev. Lett.",
    volume = "112",
    number = "13",
    pages = "132001",
    year = "2014"
}

@article{An:2025qfw,
    author = "An, Hong-Tao and Li, Yu-Shuai",
    title = "{Systematic investigation of the spectroscopy and decay behaviors of doubly-charmed pentaquarks}",
    eprint = "2512.08643",
    archivePrefix = "arXiv",
    primaryClass = "hep-ph",
    month = "12",
    year = "2025"
}

@article{Park:2023ygm,
    author = "Park, Woosung and Noh, Sungsik",
    title = "{Doubly-charmed pentaquark in a quark model with a complete set of harmonic oscillator bases}",
    eprint = "2305.00655",
    archivePrefix = "arXiv",
    primaryClass = "hep-ph",
    doi = "10.1103/PhysRevD.108.014026",
    journal = "Phys. Rev. D",
    volume = "108",
    number = "1",
    pages = "014026",
    year = "2023"
}

@article{Park:2024gbq,
    author = "Park, Woosung and Noh, Sungsik",
    title = "{Investigation on the stabilities of doubly heavy tetraquark states}",
    eprint = "2407.10443",
    archivePrefix = "arXiv",
    primaryClass = "hep-ph",
    doi = "10.1103/PhysRevD.110.074041",
    journal = "Phys. Rev. D",
    volume = "110",
    number = "7",
    pages = "074041",
    year = "2024"
}

@article{Noh:2021lqs,
    author = "Noh, Sungsik and Park, Woosung and Lee, Su Houng",
    title = "{The Doubly-heavy Tetraquarks ($qq'\bar{Q}\bar{Q'}$) in a Constituent Quark Model with a Complete Set of Harmonic Oscillator Bases}",
    eprint = "2102.09614",
    archivePrefix = "arXiv",
    primaryClass = "hep-ph",
    doi = "10.1103/PhysRevD.103.114009",
    journal = "Phys. Rev. D",
    volume = "103",
    pages = "114009",
    year = "2021"
}

@article{An:2025rjv,
    author = "An, Hong-Tao and Luo, Si-Qiang and Liu, Xiang",
    title = "{Doubly charmed hexaquarks in the diquark picture}",
    eprint = "2504.06107",
    archivePrefix = "arXiv",
    primaryClass = "hep-ph",
    doi = "10.1103/l4qq-pbr9",
    journal = "Phys. Rev. D",
    volume = "112",
    number = "5",
    pages = "054041",
    year = "2025"
}

@article{Barnes:2000hu,
    author = "Barnes, Ted and Black, N. and Swanson, E. S.",
    title = "{Meson meson scattering in the quark model: Spin dependence and exotic channels}",
    eprint = "nucl-th/0007025",
    archivePrefix = "arXiv",
    doi = "10.1103/PhysRevC.63.025204",
    journal = "Phys. Rev. C",
    volume = "63",
    pages = "025204",
    year = "2001"
}

@article{Yang:2025jsp,
    author = "Yang, Gang and Ping, Jialun and Segovia, Jorge",
    title = "{Triply heavy tetraquarks $\bar{b}c\bar{q}c$ and $\bar{c}b\bar{q}b$ in a constituent quark model}",
    eprint = "2507.13728",
    archivePrefix = "arXiv",
    primaryClass = "hep-ph",
    month = "7",
    year = "2025"
}

@article{Park:2016mez,
    author = "Park, Aaron and Park, Woosung and Lee, Su Houng",
    title = "{Dibaryons with two strange quarks and one heavy flavor in a constituent quark model}",
    eprint = "1606.01006",
    archivePrefix = "arXiv",
    primaryClass = "hep-ph",
    doi = "10.1103/PhysRevD.94.054027",
    journal = "Phys. Rev. D",
    volume = "94",
    number = "5",
    pages = "054027",
    year = "2016"
}

@article{Park:2017jbn,
    author = "Park, Woosung and Park, Aaron and Cho, Sungtae and Lee, Su Houng",
    title = "{$P_c(4380)$ in a constituent quark model}",
    eprint = "1702.00381",
    archivePrefix = "arXiv",
    primaryClass = "hep-ph",
    doi = "10.1103/PhysRevD.95.054027",
    journal = "Phys. Rev. D",
    volume = "95",
    number = "5",
    pages = "054027",
    year = "2017"
}

@article{An:2022qpt,
    author = "An, Hong-Tao and Luo, Si-Qiang and Liu, Zhan-Wei and Liu, Xiang",
    title = "{Spectroscopic behavior of fully heavy tetraquarks}",
    eprint = "2208.03899",
    archivePrefix = "arXiv",
    primaryClass = "hep-ph",
    doi = "10.1140/epjc/s10052-023-11847-7",
    journal = "Eur. Phys. J. C",
    volume = "83",
    number = "8",
    pages = "740",
    year = "2023"
}

@article{An:2022fvs,
    author = "An, Hong-Tao and Luo, Si-Qiang and Liu, Zhan-Wei and Liu, Xiang",
    title = "{Fully heavy pentaquark states in constituent quark model}",
    eprint = "2203.03448",
    archivePrefix = "arXiv",
    primaryClass = "hep-ph",
    doi = "10.1103/PhysRevD.105.074032",
    journal = "Phys. Rev. D",
    volume = "105",
    number = "7",
    pages = "074032",
    year = "2022"
}

@article{Anwar:2018sol,
    author = "Anwar, Muhammad Naeem and Ferretti, Jacopo and Santopinto, Elena",
    title = "{Spectroscopy of the hidden-charm $[qc][\bar q \bar c]$ and $[sc][\bar s \bar c]$ tetraquarks in the relativized diquark model}",
    eprint = "1805.06276",
    archivePrefix = "arXiv",
    primaryClass = "hep-ph",
    doi = "10.1103/PhysRevD.98.094015",
    journal = "Phys. Rev. D",
    volume = "98",
    number = "9",
    pages = "094015",
    year = "2018"
}

@article{Meng:2023jqk,
    author = "Meng, Lu and Chen, Yan-Ke and Ma, Yao and Zhu, Shi-Lin",
    title = "{Tetraquark bound states in constituent quark models: Benchmark test calculations}",
    eprint = "2310.13354",
    archivePrefix = "arXiv",
    primaryClass = "hep-ph",
    doi = "10.1103/PhysRevD.108.114016",
    journal = "Phys. Rev. D",
    volume = "108",
    number = "11",
    pages = "114016",
    year = "2023"
}

@article{Belle:2011aa,
    author = "Bondar, A. and others",
    collaboration = "Belle",
    title = "{Observation of two charged bottomonium-like resonances in Y(5S) decays}",
    eprint = "1110.2251",
    archivePrefix = "arXiv",
    primaryClass = "hep-ex",
    doi = "10.1103/PhysRevLett.108.122001",
    journal = "Phys. Rev. Lett.",
    volume = "108",
    pages = "122001",
    year = "2012"
}

@article{LHCb:2022sfr,
    author = "Aaij, R. and others",
    collaboration = "LHCb",
    title = "{First Observation of a Doubly Charged Tetraquark and Its Neutral Partner}",
    eprint = "2212.02716",
    archivePrefix = "arXiv",
    primaryClass = "hep-ex",
    reportNumber = "LHCb-PAPER-2022-026, CERN-EP-2022-239",
    doi = "10.1103/PhysRevLett.131.041902",
    journal = "Phys. Rev. Lett.",
    volume = "131",
    number = "4",
    pages = "041902",
    year = "2023"
}

@article{LHCb:2016axx,
    author = "Aaij, Roel and others",
    collaboration = "LHCb",
    title = "{Observation of $J/\psi\phi$ structures consistent with exotic states from amplitude analysis of $B^+\to J/\psi \phi K^+$ decays}",
    eprint = "1606.07895",
    archivePrefix = "arXiv",
    primaryClass = "hep-ex",
    reportNumber = "LHCB-PAPER-2016-018, CERN-EP-2016-155",
    doi = "10.1103/PhysRevLett.118.022003",
    journal = "Phys. Rev. Lett.",
    volume = "118",
    number = "2",
    pages = "022003",
    year = "2017"
}

@article{CMS:2013jru,
    author = "Chatrchyan, Serguei and others",
    collaboration = "CMS",
    title = "{Observation of a Peaking Structure in the $J/\psi \phi$ Mass Spectrum from $B^{\pm} \to J/\psi \phi K^{\pm}$ Decays}",
    eprint = "1309.6920",
    archivePrefix = "arXiv",
    primaryClass = "hep-ex",
    reportNumber = "CMS-BPH-11-026, CERN-PH-EP-2013-167",
    doi = "10.1016/j.physletb.2014.05.055",
    journal = "Phys. Lett. B",
    volume = "734",
    pages = "261--281",
    year = "2014"
}

@article{Liu:2024fnh,
    author = "Liu, Feng-Xiao and Ni, Ru-Hui and Zhong, Xian-Hui and Zhao, Qiang",
    title = "{Hidden and double charm-strange tetraquarks and their decays in a potential quark model}",
    eprint = "2407.19494",
    archivePrefix = "arXiv",
    primaryClass = "hep-ph",
    month = "7",
    year = "2024"
}

@article{Liang:2024met,
    author = "Liang, Zhi-Biao and Liu, Feng-Xiao and Zhong, Xian-Hui",
    title = "{All-heavy pentaquarks}",
    eprint = "2402.17974",
    archivePrefix = "arXiv",
    primaryClass = "hep-ph",
    month = "2",
    year = "2024"
}

@article{Liu:2022hbk,
    author = "Liu, Feng-Xiao and Ni, Ru-Hui and Zhong, Xian-Hui and Zhao, Qiang",
    title = "{Charmed-strange tetraquarks and their decays in the potential quark model}",
    eprint = "2211.01711",
    archivePrefix = "arXiv",
    primaryClass = "hep-ph",
    doi = "10.1103/PhysRevD.107.096020",
    journal = "Phys. Rev. D",
    volume = "107",
    number = "9",
    pages = "096020",
    year = "2023"
}

@article{LHCb:2022aki,
    author = "Aaij, Roel and others",
    collaboration = "LHCb",
    title = "{Observation of a Resonant Structure near the $D_s^+D_s^-$ Threshold in the $B^+\to D_s^+D_s^-K^+$ Decay}",
    eprint = "2210.15153",
    archivePrefix = "arXiv",
    primaryClass = "hep-ex",
    reportNumber = "CERN-EP-2022-200, LHCb-PAPER-2022-018",
    doi = "10.1103/PhysRevLett.131.071901",
    journal = "Phys. Rev. Lett.",
    volume = "131",
    number = "7",
    pages = "071901",
    year = "2023"
}

@article{LHCb:2022ogu,
    author = "Aaij, R. and others",
    collaboration = "LHCb",
    title = "{Observation of a $J/\psi\Lambda$ Resonance Consistent with a Strange Pentaquark Candidate in $B^-\to J/\psi\Lambda \bar{p}$ Decays}",
    eprint = "2210.10346",
    archivePrefix = "arXiv",
    primaryClass = "hep-ex",
    reportNumber = "CERN-EP-2022-198, LHCb-PAPER-2022-031",
    doi = "10.1103/PhysRevLett.131.031901",
    journal = "Phys. Rev. Lett.",
    volume = "131",
    number = "3",
    pages = "031901",
    year = "2023"
}

@article{Chen:2022asf,
    author = "Chen, Hua-Xing and Chen, Wei and Liu, Xiang and Liu, Yan-Rui and Zhu, Shi-Lin",
    title = "{An updated review of the new hadron states}",
    eprint = "2204.02649",
    archivePrefix = "arXiv",
    primaryClass = "hep-ph",
    doi = "10.1088/1361-6633/aca3b6",
    journal = "Rept. Prog. Phys.",
    volume = "86",
    number = "2",
    pages = "026201",
    year = "2023"
}

@article{LHCb:2021auc,
    author = "Aaij, Roel and others",
    collaboration = "LHCb",
    title = "{Study of the doubly charmed tetraquark $T_{cc}^{+}$}",
    eprint = "2109.01056",
    archivePrefix = "arXiv",
    primaryClass = "hep-ex",
    reportNumber = "CERN-EP-2021-169, LHCb-PAPER-2021-032",
    doi = "10.1038/s41467-022-30206-w",
    journal = "Nature Commun.",
    volume = "13",
    number = "1",
    pages = "3351",
    year = "2022"
}

@article{LHCb:2021vvq,
    author = "Aaij, Roel and others",
    collaboration = "LHCb",
    title = "{Observation of an exotic narrow doubly charmed tetraquark}",
    eprint = "2109.01038",
    archivePrefix = "arXiv",
    primaryClass = "hep-ex",
    reportNumber = "CERN-EP-2021-165, LHCb-PAPER-2021-031",
    doi = "10.1038/s41567-022-01614-y",
    journal = "Nature Phys.",
    volume = "18",
    number = "7",
    pages = "751--754",
    year = "2022"
}

@article{Yang:2021sue,
    author = "Yang, Xin-Dian and Wang, Fu-Lai and Liu, Zhan-Wei and Liu, Xiang",
    title = "{Newly observed $X(4630)$: a new charmoniumlike molecule}",
    eprint = "2103.03127",
    archivePrefix = "arXiv",
    primaryClass = "hep-ph",
    doi = "10.1140/epjc/s10052-021-09606-7",
    journal = "Eur. Phys. J. C",
    volume = "81",
    number = "9",
    pages = "807",
    year = "2021"
}

@article{LHCb:2021uow,
    author = "Aaij, Roel and others",
    collaboration = "LHCb",
    title = "{Observation of New Resonances Decaying to $J/\psi K^+$ and $J/\psi \phi$}",
    eprint = "2103.01803",
    archivePrefix = "arXiv",
    primaryClass = "hep-ex",
    reportNumber = "LHCb-PAPER-2020-044, CERN-EP-2021-025",
    doi = "10.1103/PhysRevLett.127.082001",
    journal = "Phys. Rev. Lett.",
    volume = "127",
    number = "8",
    pages = "082001",
    year = "2021"
}

@article{LHCb:2020jpq,
    author = "Aaij, Roel and others",
    collaboration = "LHCb",
    title = "{Evidence of a $J/\psi\Lambda$ structure and observation of excited $\Xi^-$ states in the $\Xi^-_b \to J/\psi\Lambda K^-$ decay}",
    eprint = "2012.10380",
    archivePrefix = "arXiv",
    primaryClass = "hep-ex",
    reportNumber = "LHCb-PAPER-2020-039, CERN-EP-2020-233",
    doi = "10.1016/j.scib.2021.02.030",
    journal = "Sci. Bull.",
    volume = "66",
    pages = "1278--1287",
    year = "2021"
}

@article{Wang:2020prk,
    author = "Wang, Guang-Juan and Meng, Lu and Xiao, Li-Ye and Oka, Makoto and Zhu, Shi-Lin",
    title = "{Mass spectrum and strong decays of tetraquark ${\bar{c}}{\bar{s}} qq$ states}",
    eprint = "2010.09395",
    archivePrefix = "arXiv",
    primaryClass = "hep-ph",
    doi = "10.1140/epjc/s10052-021-08978-0",
    journal = "Eur. Phys. J. C",
    volume = "81",
    number = "2",
    pages = "188",
    year = "2021"
}

@article{liu:2020eha,
    author = "liu, Ming-Sheng and Liu, Feng-Xiao and Zhong, Xian-Hui and Zhao, Qiang",
    title = "{Fully heavy tetraquark states and their evidences in LHC observations}",
    eprint = "2006.11952",
    archivePrefix = "arXiv",
    primaryClass = "hep-ph",
    doi = "10.1103/PhysRevD.109.076017",
    journal = "Phys. Rev. D",
    volume = "109",
    number = "7",
    pages = "076017",
    year = "2024"
}

@article{Xiao:2019spy,
    author = "Xiao, Li-Ye and Wang, Guang-Juan and Zhu, Shi-Lin",
    title = "{Hidden-charm strong decays of the $Z_c$ states}",
    eprint = "1912.12781",
    archivePrefix = "arXiv",
    primaryClass = "hep-ph",
    doi = "10.1103/PhysRevD.101.054001",
    journal = "Phys. Rev. D",
    volume = "101",
    number = "5",
    pages = "054001",
    year = "2020"
}

@article{Wang:2019spc,
    author = "Wang, Guang-Juan and Xiao, Li-Ye and Chen, Rui and Liu, Xiao-Hai and Liu, Xiang and Zhu, Shi-Lin",
    title = "{Probing hidden-charm decay properties of $P_c$ states in a molecular scenario}",
    eprint = "1911.09613",
    archivePrefix = "arXiv",
    primaryClass = "hep-ph",
    doi = "10.1103/PhysRevD.102.036012",
    journal = "Phys. Rev. D",
    volume = "102",
    number = "3",
    pages = "036012",
    year = "2020"
}

@article{Brambilla:2019esw,
    author = "Brambilla, Nora and Eidelman, Simon and Hanhart, Christoph and Nefediev, Alexey and Shen, Cheng-Ping and Thomas, Christopher E. and Vairo, Antonio and Yuan, Chang-Zheng",
    title = "{The $XYZ$ states: experimental and theoretical status and perspectives}",
    eprint = "1907.07583",
    archivePrefix = "arXiv",
    primaryClass = "hep-ex",
    reportNumber = "TUM-EFT 125/19",
    doi = "10.1016/j.physrep.2020.05.001",
    journal = "Phys. Rept.",
    volume = "873",
    pages = "1--154",
    year = "2020"
}

@article{Zhou:2019swr,
    author = "Zhou, Zhi-Yong and Yu, Meng-Ting and Xiao, Zhiguang",
    title = "{Decays of $X(3872)$ to $\chi_{cJ}\pi^0$ and $J/\psi\pi^+\pi^-$}",
    eprint = "1904.07509",
    archivePrefix = "arXiv",
    primaryClass = "hep-ph",
    reportNumber = "USTC-ICTS-19-08",
    doi = "10.1103/PhysRevD.100.094025",
    journal = "Phys. Rev. D",
    volume = "100",
    number = "9",
    pages = "094025",
    year = "2019"
}

@article{LHCb:2019kea,
    author = "Aaij, Roel and others",
    collaboration = "LHCb",
    title = "{Observation of a narrow pentaquark state, $P_c(4312)^+$, and of two-peak structure of the $P_c(4450)^+$}",
    eprint = "1904.03947",
    archivePrefix = "arXiv",
    primaryClass = "hep-ex",
    reportNumber = "LHCb-PAPER-2019-014 CERN-EP-2019-058",
    doi = "10.1103/PhysRevLett.122.222001",
    journal = "Phys. Rev. Lett.",
    volume = "122",
    number = "22",
    pages = "222001",
    year = "2019"
}

@article{Liu:2019zoy,
    author = "Liu, Yan-Rui and Chen, Hua-Xing and Chen, Wei and Liu, Xiang and Zhu, Shi-Lin",
    title = "{Pentaquark and Tetraquark states}",
    eprint = "1903.11976",
    archivePrefix = "arXiv",
    primaryClass = "hep-ph",
    doi = "10.1016/j.ppnp.2019.04.003",
    journal = "Prog. Part. Nucl. Phys.",
    volume = "107",
    pages = "237--320",
    year = "2019"
}

@article{Wang:2018pwi,
    author = "Wang, Guang-Juan and Liu, Xiao-Hai and Ma, Li and Liu, Xiang and Chen, Xiao-Lin and Deng, Wei-Zhen and Zhu, Shi-Lin",
    title = "{The strong decay patterns of $Z_c$ and $Z_b$ states in the relativized quark model}",
    eprint = "1811.10339",
    archivePrefix = "arXiv",
    primaryClass = "hep-ph",
    doi = "10.1140/epjc/s10052-019-7059-y",
    journal = "Eur. Phys. J. C",
    volume = "79",
    number = "7",
    pages = "567",
    year = "2019"
}

@article{Guo:2017jvc,
    author = "Guo, Feng-Kun and Hanhart, Christoph and Mei\ss{}ner, Ulf-G. and Wang, Qian and Zhao, Qiang and Zou, Bing-Song",
    title = "{Hadronic molecules}",
    eprint = "1705.00141",
    archivePrefix = "arXiv",
    primaryClass = "hep-ph",
    doi = "10.1103/RevModPhys.90.015004",
    journal = "Rev. Mod. Phys.",
    volume = "90",
    number = "1",
    pages = "015004",
    year = "2018",
    note = "[Erratum: Rev.Mod.Phys. 94, 029901 (2022)]"
}

@article{Hosaka:2016pey,
    author = "Hosaka, Atsushi and Iijima, Toru and Miyabayashi, Kenkichi and Sakai, Yoshihide and Yasui, Shigehiro",
    title = "{Exotic hadrons with heavy flavors: $X$, $Y$, $Z$, and related states}",
    eprint = "1603.09229",
    archivePrefix = "arXiv",
    primaryClass = "hep-ph",
    reportNumber = "J-PARC-TH-0046",
    doi = "10.1093/ptep/ptw045",
    journal = "PTEP",
    volume = "2016",
    number = "6",
    pages = "062C01",
    year = "2016"
}

@article{Chen:2016qju,
    author = "Chen, Hua-Xing and Chen, Wei and Liu, Xiang and Zhu, Shi-Lin",
    title = "{The hidden-charm pentaquark and tetraquark states}",
    eprint = "1601.02092",
    archivePrefix = "arXiv",
    primaryClass = "hep-ph",
    doi = "10.1016/j.physrep.2016.05.004",
    journal = "Phys. Rept.",
    volume = "639",
    pages = "1--121",
    year = "2016"
}

@article{LHCb:2015yax,
    author = "Aaij, Roel and others",
    collaboration = "LHCb",
    title = "{Observation of $J/\psi p$ Resonances Consistent with Pentaquark States in $\Lambda_b^0 \to J/\psi K^- p$ Decays}",
    eprint = "1507.03414",
    archivePrefix = "arXiv",
    primaryClass = "hep-ex",
    reportNumber = "CERN-PH-EP-2015-153, LHCB-PAPER-2015-029",
    doi = "10.1103/PhysRevLett.115.072001",
    journal = "Phys. Rev. Lett.",
    volume = "115",
    pages = "072001",
    year = "2015"
}

@article{Hiyama:2012sma,
    author = "Hiyama, Emiko",
    title = "{Gaussian expansion method for few-body systems and its applications to atomic and nuclear physics}",
    doi = "10.1093/ptep/pts015",
    journal = "PTEP",
    volume = "2012",
    pages = "01A204",
    year = "2012"
}

@article{Belle:2003nnu,
    author = "Choi, S. K. and others",
    collaboration = "Belle",
    title = "{Observation of a narrow charmonium-like state in exclusive $B^\pm \to K^\pm \pi^+ \pi^- J/\psi$ decays}",
    eprint = "hep-ex/0309032",
    archivePrefix = "arXiv",
    doi = "10.1103/PhysRevLett.91.262001",
    journal = "Phys. Rev. Lett.",
    volume = "91",
    pages = "262001",
    year = "2003"
}

@article{Hiyama:2003cu,
    author = "Hiyama, E. and Kino, Y. and Kamimura, M.",
    title = "{Gaussian expansion method for few-body systems}",
    doi = "10.1016/S0146-6410(03)90015-9",
    journal = "Prog. Part. Nucl. Phys.",
    volume = "51",
    pages = "223--307",
    year = "2003"
}

@article{Wong:2001td,
    author = "Wong, Cheuk-Yin and Swanson, E. S. and Barnes, Ted",
    title = "{Heavy quarkonium dissociation cross-sections in relativistic heavy ion collisions}",
    eprint = "nucl-th/0106067",
    archivePrefix = "arXiv",
    reportNumber = "JLAB-THY-02-86",
    doi = "10.1103/PhysRevC.66.029901",
    journal = "Phys. Rev. C",
    volume = "65",
    pages = "014903",
    year = "2002",
    note = "[Erratum: Phys.Rev.C 66, 029901 (2002)]"
}

@article{Barnes:1991em,
    author = "Barnes, Ted and Swanson, E. S.",
    title = "{A Diagrammatic approach to meson meson scattering in the nonrelativistic quark potential model}",
    reportNumber = "MIT-CTP-2026, ORNL-CCIP-91-23, UTK-91-08",
    doi = "10.1103/PhysRevD.46.131",
    journal = "Phys. Rev. D",
    volume = "46",
    pages = "131--159",
    year = "1992"
}
\end{document}